\documentclass[a4paper,11pt]{article}
\pdfoutput=1
\usepackage{jheppub}

\usepackage[utf8]{inputenc}

\usepackage{multirow}
\usepackage{textcomp,color}
\usepackage{graphicx,subfigure}

\usepackage[normalem]{ulem}

\title{Scattering of kinks of the sinh-deformed $\varphi^4$ model}

\author[a]{Dionisio Bazeia,}
\author[b]{Ekaterina Belendryasova,}
\author[b,c,1]{Vakhid A. Gani,\note{Corresponding author.}}

\affiliation[a]{Departamento de F\'\i sica, Universidade Federal da Para\'\i ba, 58051-900 Jo\~ao Pessoa, Para\'\i ba, Brazil}
\affiliation[b]{National Research Nuclear University MEPhI\\ (Moscow Engineering Physics Institute), 115409 Moscow, Russia}
\affiliation[c]{Theory Department, National Research Center Kurchatov Institute, Institute for Theoretical and Experimental Physics, 117218 Moscow, Russia}

\emailAdd{bazeia@fisica.ufpb.br}
\emailAdd{vagani@mephi.ru}

\abstract{We consider the scattering of kinks of the sinh-deformed $\varphi^4$ model, which is obtained from the well-known $\varphi^4$ model by means of the deformation procedure. Depending on the initial velocity $v_\mathrm{in}$ of the colliding kinks, different collision scenarios are realized. There is a critical value $v_\mathrm{cr}$ of the initial velocity, which separates the regime of reflection (at $v_\mathrm{in}>v_\mathrm{cr}$) and that of a complicated interaction (at $v_\mathrm{in}<v_\mathrm{cr}$) with kinks' capture and escape windows. Besides that, at $v_\mathrm{in}$ below $v_\mathrm{cr}$ we observe the formation of a bound state of two oscillons, as well as their escape at some values of $v_\mathrm{in}$.}

\makeatletter
\def\@fpheader{\relax}
\makeatother

\begin{document}

\maketitle

\flushbottom

\section{Introduction}
\label{sec:Introduction}

Topological defects arise in a diversity of contexts in high energy physics, cosmology, quantum and classical field theory, condensed matter, and so on. In high energy physics, they are topologically non-trivial solutions of the equations of motion and possess very interesting properties, which lead to new physical phenomena \cite{Rajaraman.book.1982,Vilenkin.book.2000,Manton.book.2004,Vachaspati.book.2006}.

Nowadays, the study of the topological defects is a very fast developing area with significant effort being applied to the investigation of domain walls, vortices, strings, as well as embedded topological defects such as a Q-lump on a domain wall and a skyrmion on a domain wall, and so on \cite{nitta3,nitta4,jennings,nitta5,blyankinshtein,Loginov.YadFiz.2011.eng,Loginov.YadFiz.2011.rus,GaKsKu01.eng,GaKsKu01.rus,GaKsKu02.eng,GaKsKu02.rus,Bazeia.EPJC.2017,Bazeia.fermion.2017}. It is also of interest to mention the so-called Q-balls and similar configurations \cite{Schweitzer.NPA.2016,Nugaev.PRD.2013,Bazeia.EPJC.2016,Bazeia.PLB.2016,Bazeia.PLB.2017,Dzhunushaliev.PRD.2016}, which are charged and protected against decaying into the elementary excitations supported by the respective model. Also, it is worth mentioning other possibilities, such as the study of solitons in fibers \cite{kivshar}, bubble collisions in cosmology \cite{bubble}, and localized excitations in nonlinear systems \cite{malomed}.

Models in $(1,1)$ space-time dimensions are of special interest \cite{Vilenkin.book.2000,Manton.book.2004,Kudryavtsev.UFN.1997.eng,Kudryavtsev.UFN.1997.rus}, since the dynamics of some two- or three-dimensional systems can be reduced to the one-dimensional models. For example, a planar domain wall, which separates regions with different minima of the potential, in the direction perpendicular to it can be interpreted as a one-dimensional topological configuration (a kink). On the other hand, the $(1,1)$-dimensional field-theoretical models can be a first step towards more complicated higher-dimensional models. Moreover, even in the $(1,1)$-dimensional case, topological defects may arise in more complex models with two or more fields, see, e.g., refs.~\cite{Rajaraman.PRL.1979,Ruck.NPB.1980,MacKenzie.NPB.1988,Abraham.NPB.1991,Bazeia.PLA.1995,Morris.PRD.1995,Bazeia.PLA.1996,Bazeia.PRE.1996,Chibisov.PRD.1997,Morris.IJMPA.1998,Edelstein.PRD.1998,Bazeia.JHEP.1999,Gibbons.PRL.1999,Oda.PLB.1999,Carroll.PRD.2000,Bazeia.PRL.2000,Bazeia.PRD.2000,Lensky.JETP.2001.eng,Lensky.JETP.2001.rus,Bazeia.MPLA.2002,Alonso_Izquierdo.PRD.2002,Kurochkin.CMMP.2004.eng,Kurochkin.CMMP.2004.rus,Bazeia.AHEP.2013,Bazeia.EPJC.2014,Katsura.PRD.2014,GaLiRa,GaLiRaconf,Akula.EPJC.2016,GaKiRu,GaKiRu.conf}. In this more general context, several works have developed analytical solutions, which, in turn, has allowed one to study their stability and to use them in application of interest in physics   \cite{Rajaraman.PRL.1979,Ruck.NPB.1980,MacKenzie.NPB.1988,Abraham.NPB.1991,Bazeia.PLA.1995,Morris.PRD.1995,Bazeia.PLA.1996,Bazeia.PRE.1996,Chibisov.PRD.1997,Morris.IJMPA.1998,Edelstein.PRD.1998,Bazeia.JHEP.1999}. Other investigations have dealt with the presence of junctions and/or intersections of defects \cite{Gibbons.PRL.1999,Oda.PLB.1999,Carroll.PRD.2000,Bazeia.PRL.2000}, and with issues related to composite-kink internal structures, twinlike models with several fields and scalar triplet on domain walls \cite{Bazeia.PRD.2000,Lensky.JETP.2001.eng,Lensky.JETP.2001.rus,Bazeia.MPLA.2002,Alonso_Izquierdo.PRD.2002,Kurochkin.CMMP.2004.eng,Kurochkin.CMMP.2004.rus,Bazeia.AHEP.2013,Bazeia.EPJC.2014,Katsura.PRD.2014,GaLiRa,GaLiRaconf,Akula.EPJC.2016}, among other issues.

In the case of models described by real scalar fields with standard kinematics, in the $(1,1)$ space-time the presence of interactions that develop spontaneous symmetry breaking in general leads to localized topological structures having the kinklike profile. The interactions of these one-dimensional topological structures with each other and with spatial inhomogeneities (impurities) have attracted the attention of physicists and mathematicians for a long time; see, e.g., refs.~\cite{Kudryavtsev.UFN.1997.eng,Kudryavtsev.UFN.1997.rus}. The first studies on this subject date back to the 1970s and 1980s \cite{Kudryavtsev.JETPLett.1975.eng,Kudryavtsev.JETPLett.1975.rus,Campbell.phi4.1983}. Nevertheless, forty years later we see that it is still an actively developing area with many new applications. Many important results have been obtained by means of the numerical simulation, which is one of the most powerful tools for studying the subject. In particular, resonance phenomena -- escape windows and quasi-resonances -- were found in the kink-antikink scattering process. A broad class of $(1,1)$-dimensional models with polynomial potentials such as the $\varphi^4$, $\varphi^6$, $\varphi^8$ models, and those with higher degree polynomial self-interaction has been considered \cite{Campbell.phi4.1983,lohe,GaKuLi,MGSDJ,dorey,khare,GaLeLi,GaLeLiconf,Weigel.conf.2017,Weigel.PLB.2017,Dorey.JHEP.2017,Bazeia.PRD.2006,Radomskiy.JPCS.2017,Belendryasova.arXiv.2017,Belendryasova.conf.2017}. One should also mention the new results on the long-range interaction between kinks \cite{Belendryasova.arXiv.2017,Belendryasova.conf.2017,Radomskiy.JPCS.2017,Guerrero.PRE.1997,Guerrero.PLA.1998,Guerrero.Phys_A.1998}. Other models with non-polynomial potentials are also being discussed in the literature. For example, the modified sine-Gordon \cite{Peyrard.msG.1983}, the double sine-Gordon \cite{Campbell.1986,Campbell.dsG.1986,GaKuPRE}, and a variety of models which can be obtained using the deformation procedure, which we explain below.

Apart from the numerical solving of the equation of motion, other methods are widely used for investigating the kink-antikink interactions. One of them is the collective coordinate method \cite{GaKuLi,Weigel.cc.2014,Weigel.cc.2016,Demirkaya.cc.2017,Baron.cc.2014,Javidan.cc.2010,Christov.cc.2008,GaKu.SuSy.2001.eng,GaKu.SuSy.2001.rus}. Within this approximation a real field-theoretical system (which formally has an infinite number of degrees of freedom) is approximately described as a system with one or a few degrees of freedom. For example, in the case of the kink-antikink configuration one can use the distance between the kink and the antikink as the only degree of freedom (collective coordinate). In more complicated modifications of this approach other degrees of freedom (for instance, vibrational ones) can be involved, see, e.g., \cite{Weigel.cc.2014,Weigel.cc.2016,Demirkaya.cc.2017}. Another approximation, which allows to estimate the force between kink and antikink, is the Manton's method \cite[Ch.~5]{Manton.book.2004}, \cite{perring62,rajaram77,Manton.npb.1979,KKS.PRE.2004}. This method is based on using the kinks' asymptotics in situations where the distance between the kinks is large. However, one should mention that the applicability of this method for kinks and solitons with power-law asymptotics is not obvious.

An impressive progress has been achieved in the analytical treatment of the $(1,1)$-dimensional field-theoretical models. Among several possibilities to deal with the problem analytically, the trial orbit method was suggested in \cite{Rajaraman.PRL.1979} as a way to solve the equations of motion in systems described by two real scalar fields that interact nonlinearly. This method has been used by others, and in \cite{Bazeia.MPLA.2002} it was shown to be very effective when the equations of motion can be reduced to first-order differential equations. Also, in \cite{Alonso_Izquierdo.PRD.2002} the authors have used the integrating factor to solve the equations of motion in the case of a very specific potential.

Another possibility of searching for models that support analytical solutions appeared before in \cite{Bazeia.PRD.2002} and also in refs.~\cite{Bazeia.PRD.2004,Bazeia.PRD.2006.braneworlds}. It refers to the deformation procedure, a method of current interest which helps us to introduce new models, and solve them analytically. This will be further reviewed below, and used to define the model \cite{Bazeia.arXiv.2017.sinh.conf} we want to investigate in the current work. In particular, the new model is somehow similar to the $\varphi^4$ model with spontaneous symmetry breaking, so we will compare its features with the $\varphi^4$ case, in order to highlight the differences between the two cases, and to see how the non-polynomial interaction of the new model modifies the behavior seen in the standard $\varphi^4$ model.

In this work we focus our attention on the kink-antikink scattering process and organize the investigation as follows. In section \ref{sec:Defects} we give general introduction to the $(1,1)$-dimensional field-theoretical models, which possess topological solutions with the kink profile. In section \ref{sec:phi-4} we review the $\varphi^4$ model, briefly accounting for the kink-antikink scattering within this model. Furthermore, in section~\ref{sec:sinh-phi-4} we apply the deformation procedure to the $\varphi^4$ model in order to introduce a model with non-polynomial potential, which we call the sinh-deformed $\varphi^4$ model. In section \ref{sec:Scattering_sinh_phi-4} we focus on the collisions of the kink and the antikink of the sinh-deformed $\varphi^4$ model. In this section we present our main results and compare them with the results of the $\varphi^4$ model. Finally, in section \ref{sec:Conclusion} we conclude with a discussion of the results and the prospects for future works.


\section{Topological solitons in (1,1)-dimensional models}
\label{sec:Defects}

Consider a field-theoretical model in the $(1,1)$-dimensional space-time with its dynamics defined by the Lagrangian
\begin{equation}\label{eq:Lagrangian}
\mathcal{L}=\frac{1}{2}\left(\frac{\partial\varphi}{\partial t}\right)^2-\frac{1}{2}\left(\frac{\partial\varphi}{\partial x}\right)^2-U(\varphi),
\end{equation}
where $\varphi(x,t)$ is a real scalar field. The potential $U(\varphi)$ is supposed to be non-negative function with two or more degenerate minima, $\varphi_1^{(0)}$, $\varphi_2^{(0)}$, \dots, such that $U(\varphi_1^{(0)})=U(\varphi_2^{(0)})=...=0$. The Lagrangian \eqref{eq:Lagrangian} leads to the following equation of motion for the field $\varphi$
\begin{equation}\label{eq:eqmo}
\frac{\partial^2\varphi}{\partial t^2} - \frac{\partial^2\varphi}{\partial x^2} + \frac{dU}{d\varphi}=0.
\end{equation}
The energy functional corresponding to the Lagrangian \eqref{eq:Lagrangian} is
\begin{equation}\label{eq:energy}
E[\varphi] = \int_{-\infty}^{\infty}\left[\frac{1}{2} \left( \frac{\partial\varphi}{\partial t} \right)^2 + \frac{1}{2} \left( \frac{\partial\varphi}{\partial x} \right) ^2 + U(\varphi)\right]dx.
\end{equation}
In the static case $\displaystyle\frac{\partial\varphi}{\partial t}=0$, and from eq.~\eqref{eq:eqmo} we have
\begin{equation}\label{eq:eqmo_static}
\frac{d^2\varphi}{dx^2}=\frac{dU}{d\varphi}.
\end{equation}
This equation can be easily transformed into the first order differential equations
\begin{equation}\label{eq:eqmo_BPS}
\frac{d\varphi}{dx} = \pm\sqrt{2U}.
\end{equation}
For the energy of the static configuration to be finite, the two following conditions must hold
\begin{equation}\label{eq:plus_minus_infty}
\lim_{x\to-\infty}\varphi(x) = \varphi_i^{(0)} \quad\mbox{and}\quad
\lim_{x\to+\infty}\varphi(x) = \varphi_j^{(0)},
\end{equation}
where $\varphi^{(0)}_i$ and $\varphi^{(0)}_j$ are two adjacent minima of the potential. These expressions \eqref{eq:plus_minus_infty} are necessary conditions for the energy of a static configuration to be finite. If \eqref{eq:plus_minus_infty} hold, then the second and the third terms in the integrand in \eqref{eq:energy} fall off at $x\to\pm\infty$, hence the integral \eqref{eq:energy} can be convergent. Configurations  with $\varphi_i^{(0)}\neq\varphi_j^{(0)}$ are called topological and have a kinklike shape. In this sense, a conserved topological current can be introduced, and for the models to be investigated below one can use
\begin{equation}\label{eq:top_current}
j_\mathrm{top}^\mu = \frac{1}{2}\varepsilon^{\mu\nu}\partial_\nu\varphi.
\end{equation}
The corresponding topological charge is
\begin{equation}\label{eq:top_charge}
Q_\mathrm{top}^{} = \int_{-\infty}^{\infty}j_\mathrm{top}^0dx = \frac{1}{2} \left[ \varphi(+\infty)-\varphi(-\infty) \right].
\end{equation}
This charge is determined only by the asymptotics \eqref{eq:plus_minus_infty}, so it does not depend on the behavior of the field $\varphi(x)$ at finite $x$.

For every non-negative potential $U(\varphi)$ we can introduce a smooth function $W(\varphi)$, called the superpotential, as
\begin{equation}\label{eq:dwdfi}
U(\varphi) = \frac{1}{2}\left(\frac{dW}{d\varphi}\right)^2.
\end{equation}
Using the superpotential we can rewrite the energy of a static configuration $\varphi(x)$ in the following manner
\begin{equation}\label{eq:energy_with_BPS}
E = E_\mathrm{BPS}^{} + \frac{1}{2}\int_{-\infty}^{\infty}\left(\frac{d\varphi}{dx}\pm\frac{dW}{d\varphi}\right)^2dx,
\end{equation}
where
\begin{equation}\label{eq:energy_BPS}
E_\mathrm{BPS}^{} = |W[\varphi(+\infty)]-W[\varphi(-\infty)]|.
\end{equation}
From eq.~\eqref{eq:energy_with_BPS} one can see that the energy of any static configuration belonging to a given topological sector is bounded from below by $E_\mathrm{BPS}^{}$. The configurations with the minimal energy \eqref{eq:energy_BPS} are called BPS configurations, or BPS saturated configurations \cite{bps1.eng,bps1.rus,bps2}. From eq.~\eqref{eq:energy_with_BPS} it is easy to see that any BPS configuration satisfies the first order differential equations
\begin{equation}\label{eq:eqmo_with_superpotential}
\frac{d\varphi}{dx} = \pm\frac{dW}{d\varphi},
\end{equation}
which coincide with \eqref{eq:eqmo_BPS}.

Below we deal with kinks and antikinks --- the BPS saturated topological solutions of eq.~\eqref{eq:eqmo_BPS}, which interpolate between neighboring minima of the potential. The solution with the asymptitics $\varphi(+\infty)>\varphi(-\infty)$ is called kink, while the term antikink stands for the solution with $\varphi(+\infty)<\varphi(-\infty)$. Sometimes we use the term kink for both kink and antikink, for brevity.

Many phenomena observed in the kink-antikink scattering can be explained by the presence of the vibrational mode(s) in the kink's excitation spectrum. In order to find the spectrum of localized excitations of a kink, we have to add a small perturbation $\delta\varphi(x,t)$ to the static kink solution $\varphi_\mathrm{k}^{}(x)$,
\begin{equation}
\varphi(x,t) = \varphi_\mathrm{k}^{}(x) + \delta\varphi(x,t), \quad |\delta\varphi| \ll |\varphi_\mathrm{k}^{}|.
\end{equation}
The substitution of $\varphi(x,t)$ into the equation of motion \eqref{eq:eqmo} leads to the partial differential equation for the perturbation $\delta\varphi(x,t)$; after linearization one gets
\begin{equation}\label{eq:eq_for_delta_phi}
\frac{\partial^2\delta\varphi}{\partial t^2} - \frac{\partial^2\delta\varphi}{\partial x^2} + \left.\frac{d^2U}{d\varphi^2}\right|_{\varphi_\mathrm{k}^{}(x)}\,\delta\varphi = 0.
\end{equation}
Since the second derivative of the potential calculated at the static solution $\varphi_\mathrm{k}^{}(x)$ depends only on $x$, we can assume that $\delta\varphi$ has the form
\begin{equation}\label{eq:delta_phi}
\delta\varphi(x,t) = \eta(x)\cos\:\omega t,
\end{equation}
and this allows us to obtain the eigenvalue problem of the type of the stationary Schr\"odinger equation,
\begin{equation}\label{eq:Schrodinger}
\hat{H}\eta(x) = \omega^2\eta(x),
\end{equation}
where the operator $\hat{H}$ (the Hamiltonian) is
\begin{equation}\label{eq:Schrod_Ham}
\hat{H} = -\frac{d^2}{dx^2} + u(x),
\end{equation}
with the potential
\begin{equation}\label{eq:Schrod_potential}
u(x) = \left.\frac{d^2U}{d\varphi^2}\right|_{\varphi_\mathrm{k}^{}(x)}.
\end{equation}
For each state of the discrete spectrum, the corresponding eigenfunction $\eta(x)$ is a twice continuously differentiable and square-integrable on the $x$-axis. Kink and antikink have the same excitation spectrum.

The discrete spectrum in the potential \eqref{eq:Schrod_potential} always possesses a zero (or translational) mode $\omega_0^{}=0$. It can easily be shown by differentiating eq.~\eqref{eq:eqmo_static} with respect to $x$, and taking into account that $\varphi_\mathrm{k}^{}(x)$ is a solution of eq.~\eqref{eq:eqmo_static}, i.e.
\begin{equation}
-\frac{d^2}{dx^2}\frac{d\varphi_\mathrm{k}^{}}{dx} + \left.\frac{d^2U}{d\varphi^2}\right|_{\varphi_\mathrm{k}^{}(x)}\,\frac{d\varphi_\mathrm{k}^{}}{dx} = 0,
\quad\mbox{or}\quad
\hat{H}\cdot\frac{d\varphi_\mathrm{k}^{}}{dx} = 0.
\end{equation}
So we see that $\displaystyle\frac{d\varphi_\mathrm{k}^{}}{dx}$ is an eigenfunction of the Hamiltonian \eqref{eq:Schrod_Ham} associated with the eigenvalue $\omega_0^{}=0$. The presence of a zero mode in the kink's excitation spectrum is a consequence of the translational invariance of the Lagrangian.

Furthermore, the presence of $W$ in eq.~\eqref{eq:dwdfi} allows to write the operator $\hat H$ in the form
\begin{equation}
{\hat H}=A^\dag A,
\end{equation}
where $A^\dag$ and $A$ are the first order differential operators
\begin{equation}
A^\dag=\frac{d}{dx}+\left.\frac{d^2W}{d\varphi^2}\right|_{\varphi_\mathrm{k}^{}(x)},\quad
A=-\frac{d}{dx}+\left.\frac{d^2W}{d\varphi^2}\right|_{\varphi_\mathrm{k}^{}(x)}.
\end{equation}
This factorization shows that the operator $\hat H$ is non-negative, so the static solution $\varphi_{\mathrm{k}}(x)$ is linearly stable.


\section{The $\varphi^4$ model}
\label{sec:phi-4}

\begin{figure}[h!]
\centering
\begin{minipage}{0.47\linewidth}
\subfigure[\:Potentials]{
\includegraphics[width=\linewidth]{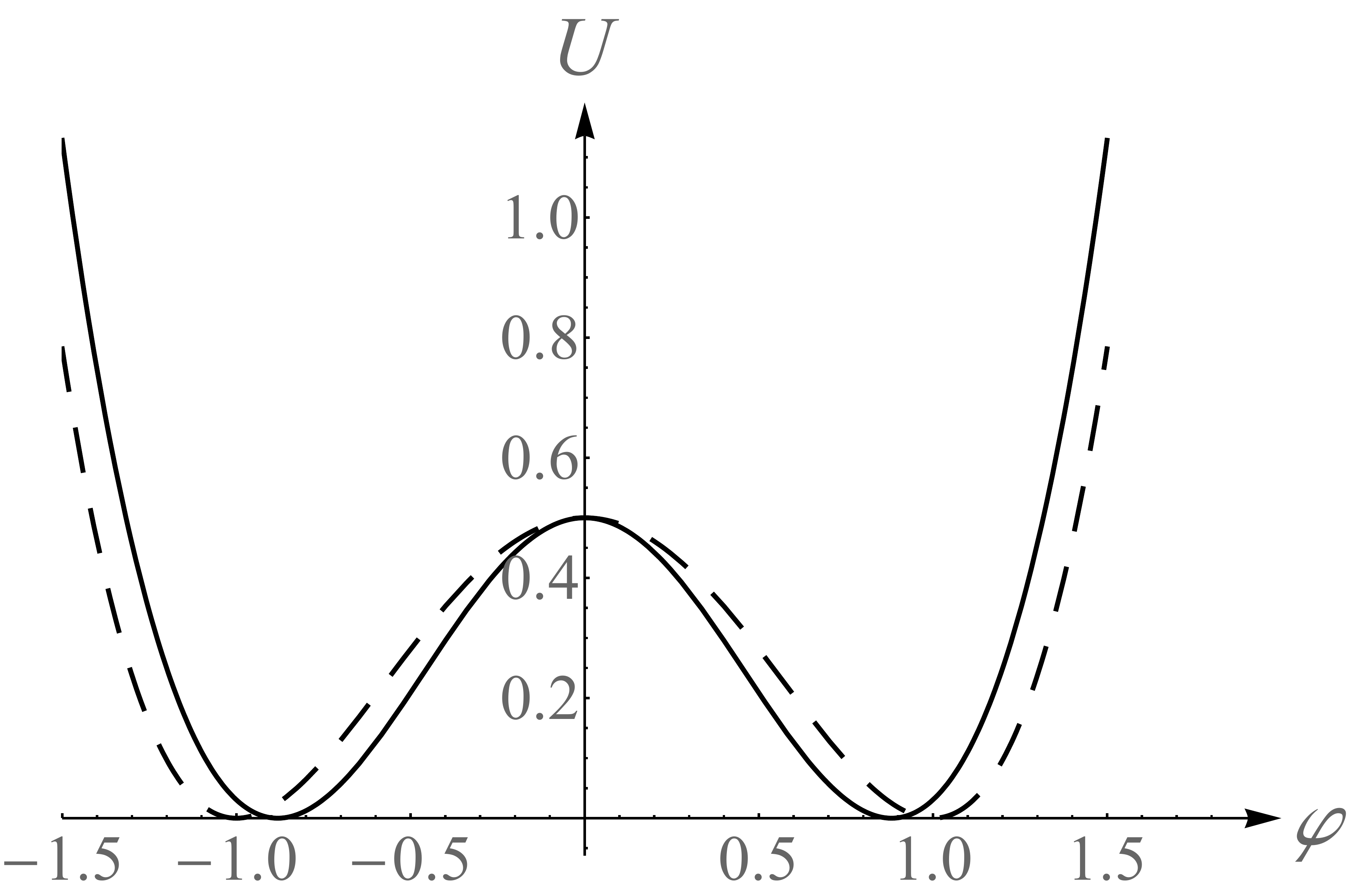}
\label{fig:potentials}
}
\end{minipage}
\hfill
\begin{minipage}{0.47\linewidth}
\subfigure[\:Kinks]{
\includegraphics[width=\linewidth]{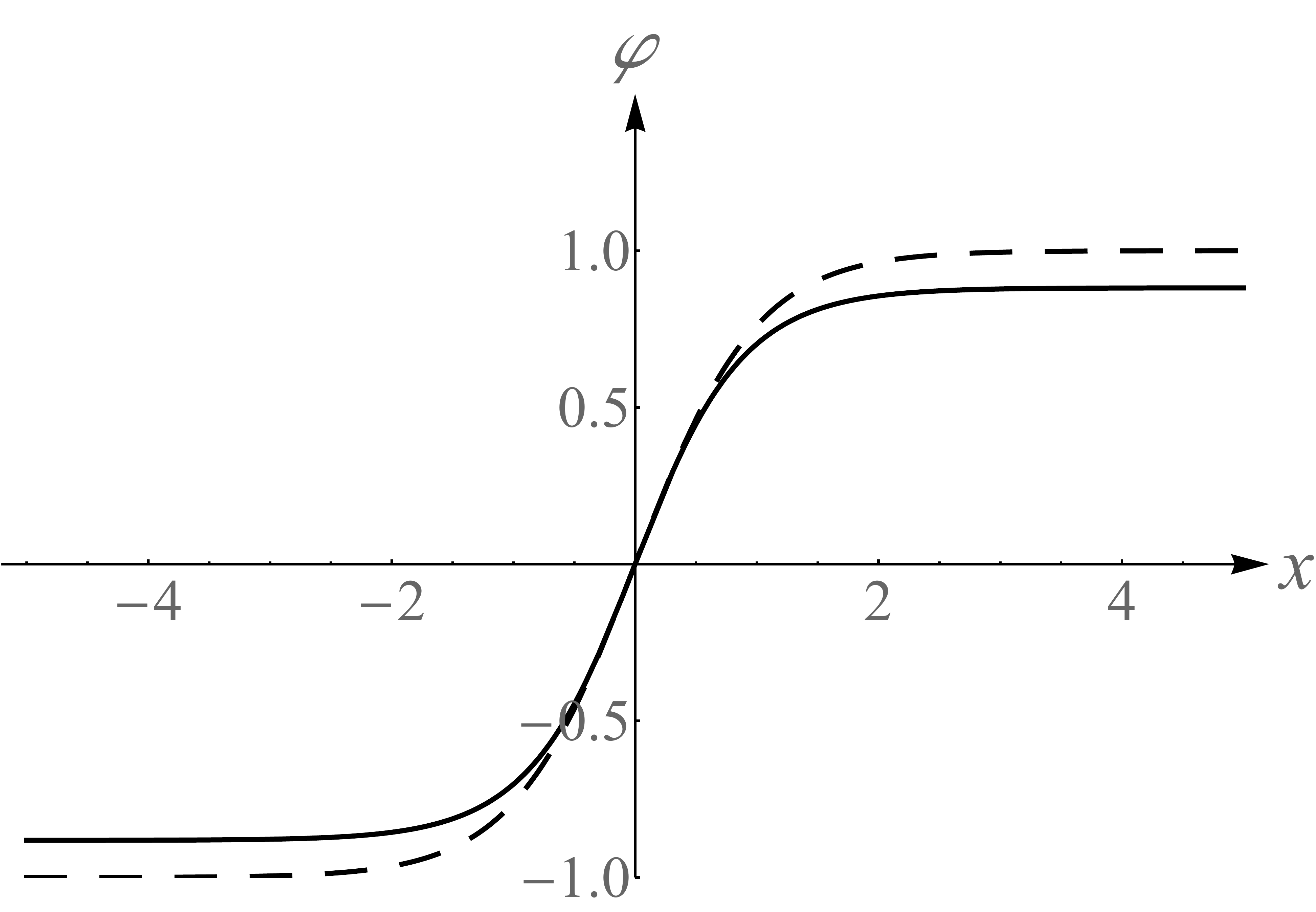}
\label{fig:kinks}
}
\end{minipage}
\caption{Potentials and kinks of the $\varphi^4$ model (dashed curves) and of the sinh-deformed $\varphi^4$ model (solid curves).}
\label{fig:potentials_and_kinks}
\end{figure}

In this section we recall some facts about kinks of the $\varphi^4$ model. We use for simplicity dimensionless fields and space-time coordinates and write the potential of the $\varphi^4$ model in the form
\begin{equation}\label{eq:potential_phi-4}
U_1^{}(\varphi)=\frac{1}{2}(1-\varphi^2)^2.
\end{equation}
This potential possesses two degenerate minima $\varphi_1^{(0)}=-1$ and $\varphi_2^{(0)}=1$, see figure \ref{fig:potentials}.
The equation \eqref{eq:eqmo_BPS} with the potential \eqref{eq:potential_phi-4} can be easily integrated, which yields the static topologically non-trivial solutions, the kink and the antikink:
\begin{equation}\label{eq:kinks_phi-4}
\varphi_\mathrm{k}^{}(x)=\tanh x,\quad \varphi_\mathrm{\bar{k}}^{}(x)=-\tanh x.
\end{equation}
These kinks interpolate between the two minima of the potential, as shown in figure \ref{fig:kinks}. The mass of the kink (antikink), i.e.\ the energy $E[\varphi_\mathrm{k}^{}(x)]$ of the static kink (antikink), is
\begin{equation}
M_\mathrm{k} = \frac{4}{3}.
\end{equation}
The moving kink (antikink) can be obtained from eq.~\eqref{eq:kinks_phi-4} by the Lorentz boost.

The quantum-mechanical potential \eqref{eq:Schrod_potential}, which defines the spectrum of the localized excitations of the $\varphi^4$ kink, has the form
\begin{equation}\label{eq:Schrod_potential_phi4}
u_1^{}(x) = 4 - \frac{6}{\cosh^2x}.
\end{equation}
It is the well-known modified P\"oschl-Teller potential \cite{PT}. Apart from the zero mode $\omega_0^{}=0$, there is a vibrational mode with the frequency $\omega_1^{}=\sqrt{3}$. As we explain below, the presence of the vibrational mode leads to resonance phenomena in the kink-antikink collisions.

As we informed in the Introduction, the scattering of the $\varphi^4$ kinks is well-studied, so let us now briefly review the main features of the collision processes in this case.

Consider the initial configuration in the form of kink and antikink centered at the points $x=-x_0^{}$ and $x=x_0^{}$, respectively, and moving towards each other with the initial velocities $v_\mathrm{in}^{}$ in the laboratory frame, i.e.
\begin{equation}\label{eq:in_cond_phi-4}
\varphi(x,t) = \tanh\left(\frac{x+x_0^{}-v_\mathrm{in}^{}t }{\sqrt{1-v_\mathrm{in}^2}}\right) - \tanh\left(\frac{x-x_0^{}+v_\mathrm{in}^{}t}{\sqrt{1-v_\mathrm{in}^2}}\right) - 1.
\end{equation}
To find evolution of this initial configuration, we solved the equation of motion \eqref{eq:eqmo} with the potential \eqref{eq:potential_phi-4} numerically using the standard explicit finite difference scheme,
\begin{equation}\label{eq:difference_scheme}
\frac{\partial^2\varphi}{\partial t^2} = \frac{\varphi_i^{j+1}-2\varphi_i^j+\varphi_i^{j-1}}{\delta t^2}, \quad \frac{\partial^2\varphi}{\partial x^2} = \frac{\varphi_{i+1}^j-2\phi_i^j+\varphi_{i-1}^j}{\delta x^2},
\end{equation}
where $(i,j)$ number the $x$ and $t$ coordinates of the grid points, $(x_i,t_j)$, on a grid with the steps $\delta t=0.008$ and $\delta x=0.01$. We repeated selected computations with smaller steps, $\delta t=0.004$ and $\delta x=0.005$, in order to check our numerical results. We also checked the total energy conservation. In all simulations of the $\varphi^4$ kinks collisions we used the initial half-distance $x_0^{}=5$.

\begin{figure}[h!]
\begin{minipage}{0.47\linewidth}
\centering\includegraphics[width=\linewidth]{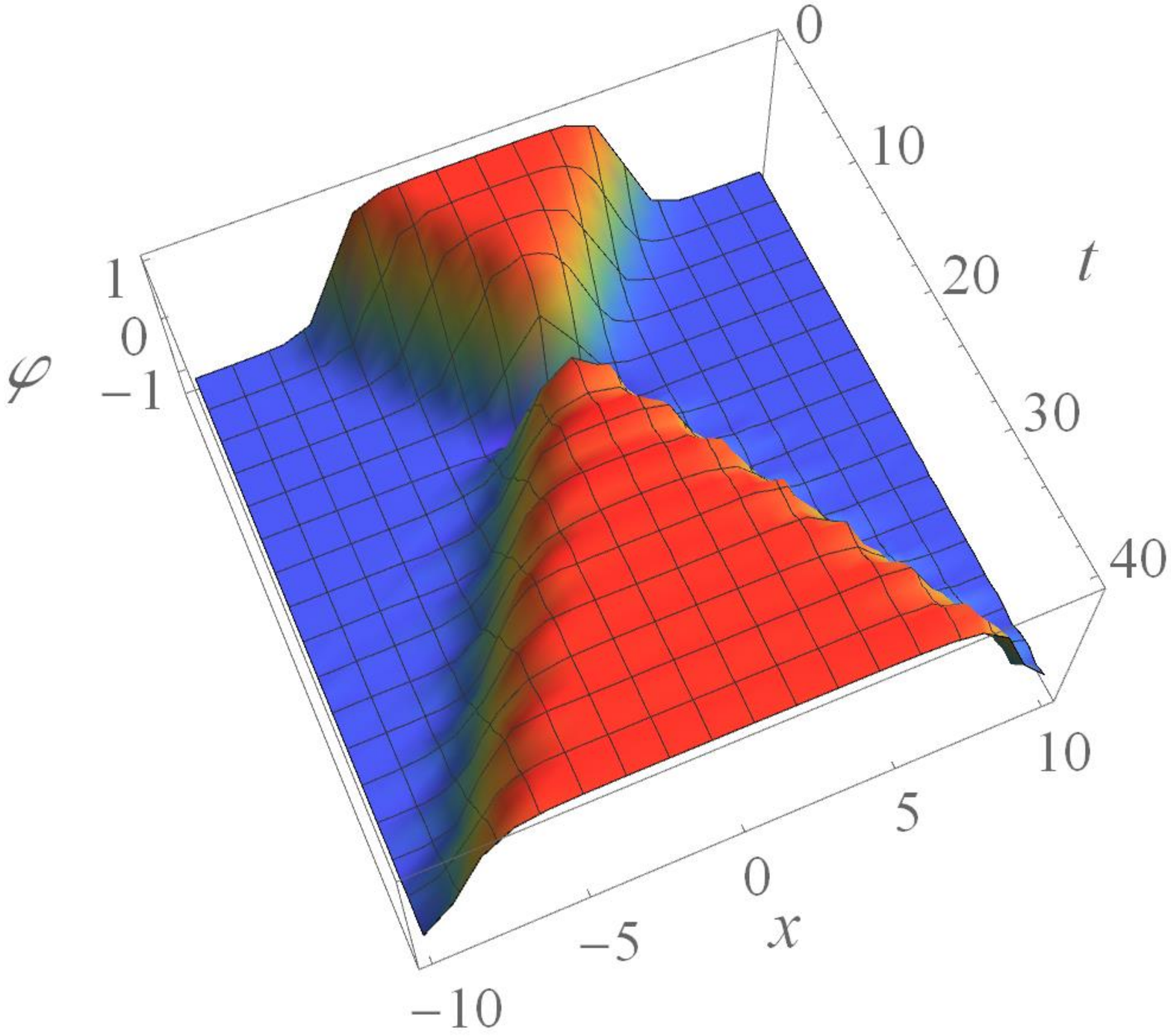}
\end{minipage}
\hspace{5mm}
\begin{minipage}{0.47\linewidth}
\centering\includegraphics[width=\linewidth]{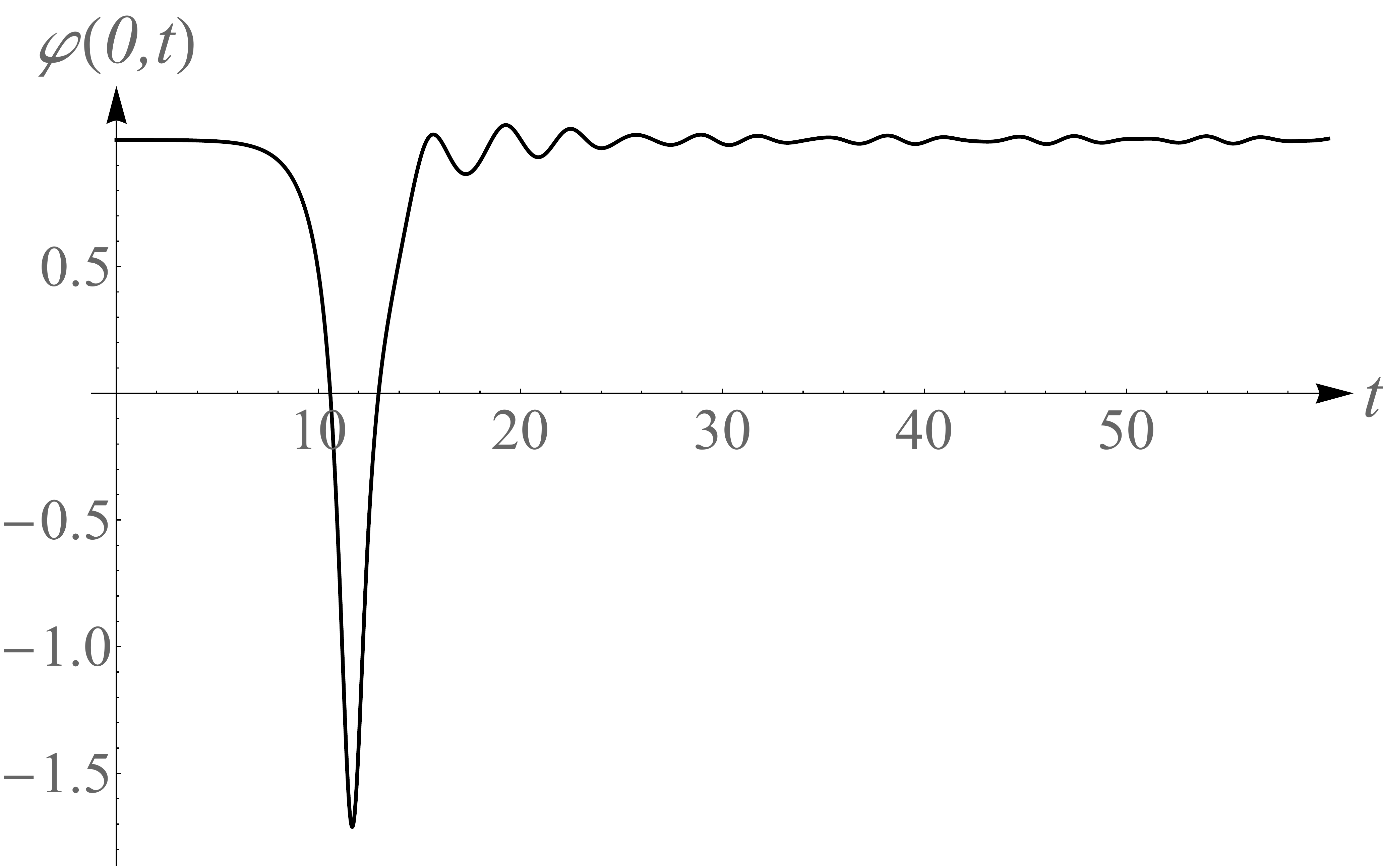}
\end{minipage}
\caption{Bounce off and escape of kinks of the $\varphi^4$ model at $v_\mathrm{in}=0.4000$. {\bf Left panel} --- the space-time picture of the field evolution. {\bf Right panel} --- the time dependence of the field at the origin.}
\label{fig:escape_phi-4}
\end{figure}

Depending on the initial velocity, the kinks scattering looks differently. There is a critical value of the initial velocity $v_\mathrm{cr}^{}\approx 0.2598$. At $v_\mathrm{in}^{}>v_\mathrm{cr}^{}$ we observe kinks escape after a collision, see figure \ref{fig:escape_phi-4}.
Some part of the energy is being emitted in the form of small waves. 

\begin{figure}[h!]
\begin{minipage}{0.47\linewidth}
\centering\includegraphics[width=\linewidth]{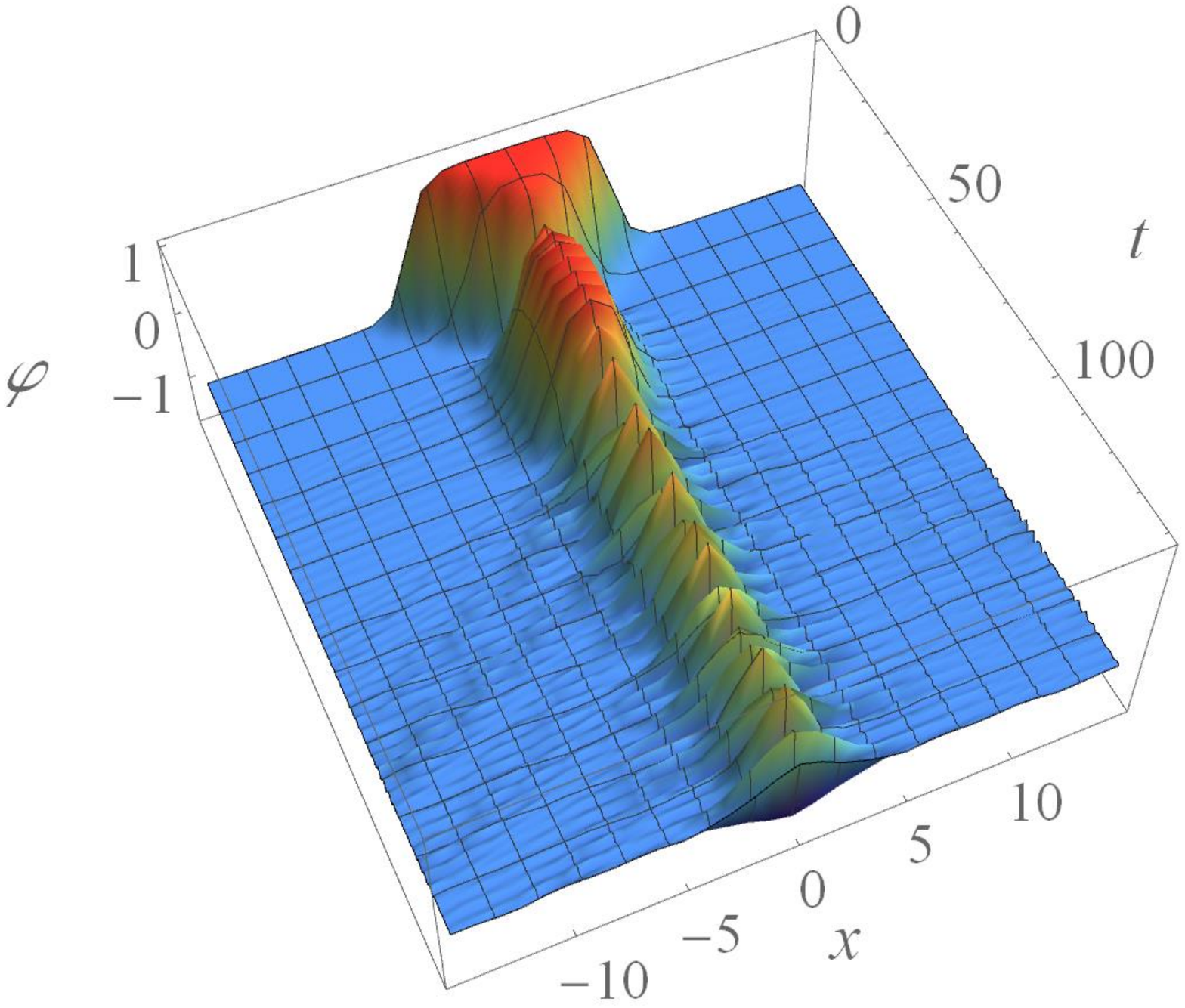}
\end{minipage}
\hspace{5mm}
\begin{minipage}{0.47\linewidth}
\centering\includegraphics[width=\linewidth]{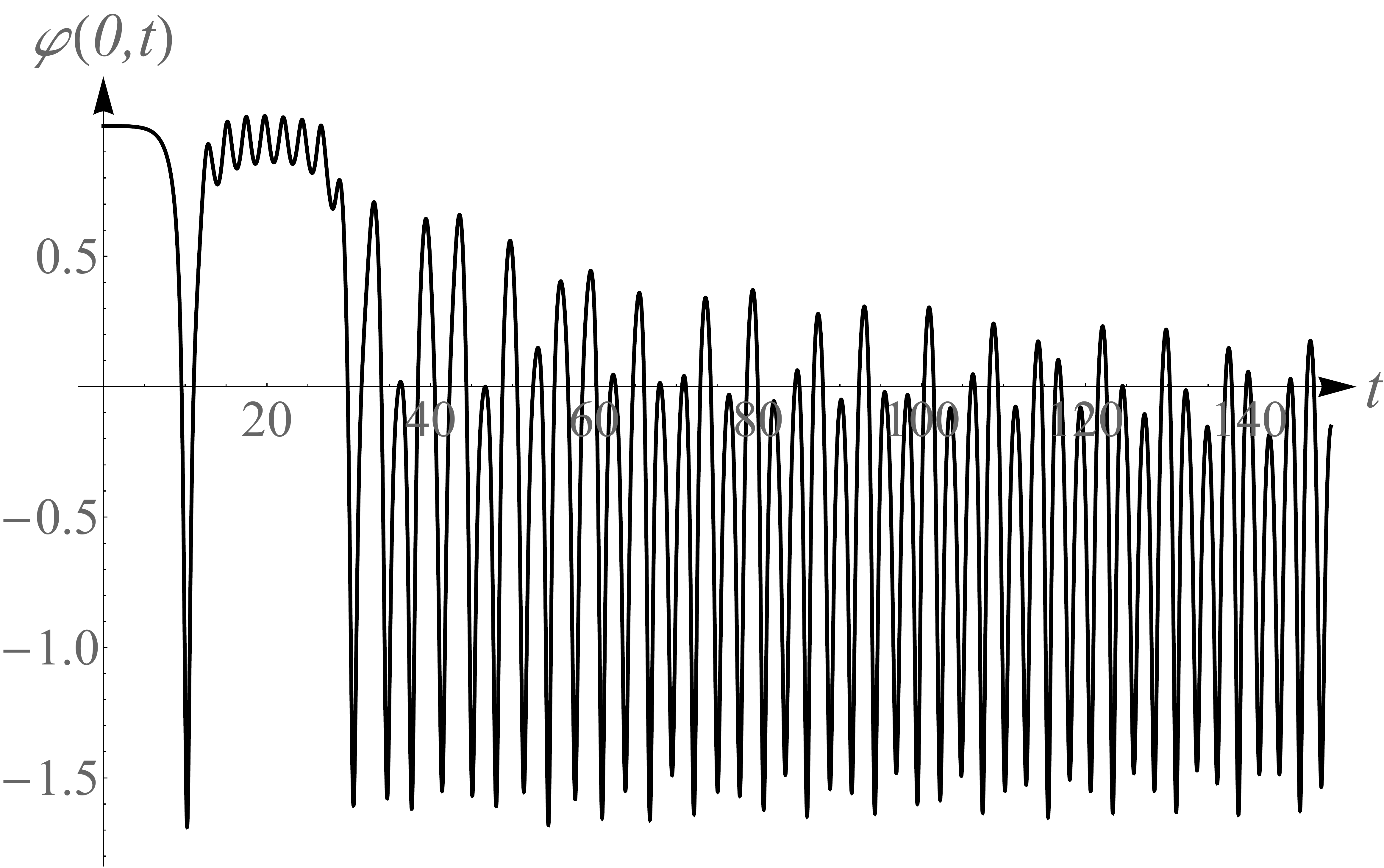}
\end{minipage}
\caption{The formation of a bound state in the $\varphi^4$ kink-antikink collision at $v_\mathrm{in}^{}=0.2541$. {\bf Left panel} --- the space-time picture of a bion formation. {\bf Right panel} --- the time dependence of the field at the origin.}
\label{fig:bion_phi-4}
\end{figure}

\begin{figure}[h!]
\subfigure[\ Two-bounce window, $v_\mathrm{in}^{}=0.2528$]{
\begin{minipage}{0.47\linewidth}
\centering\includegraphics[width=\linewidth]{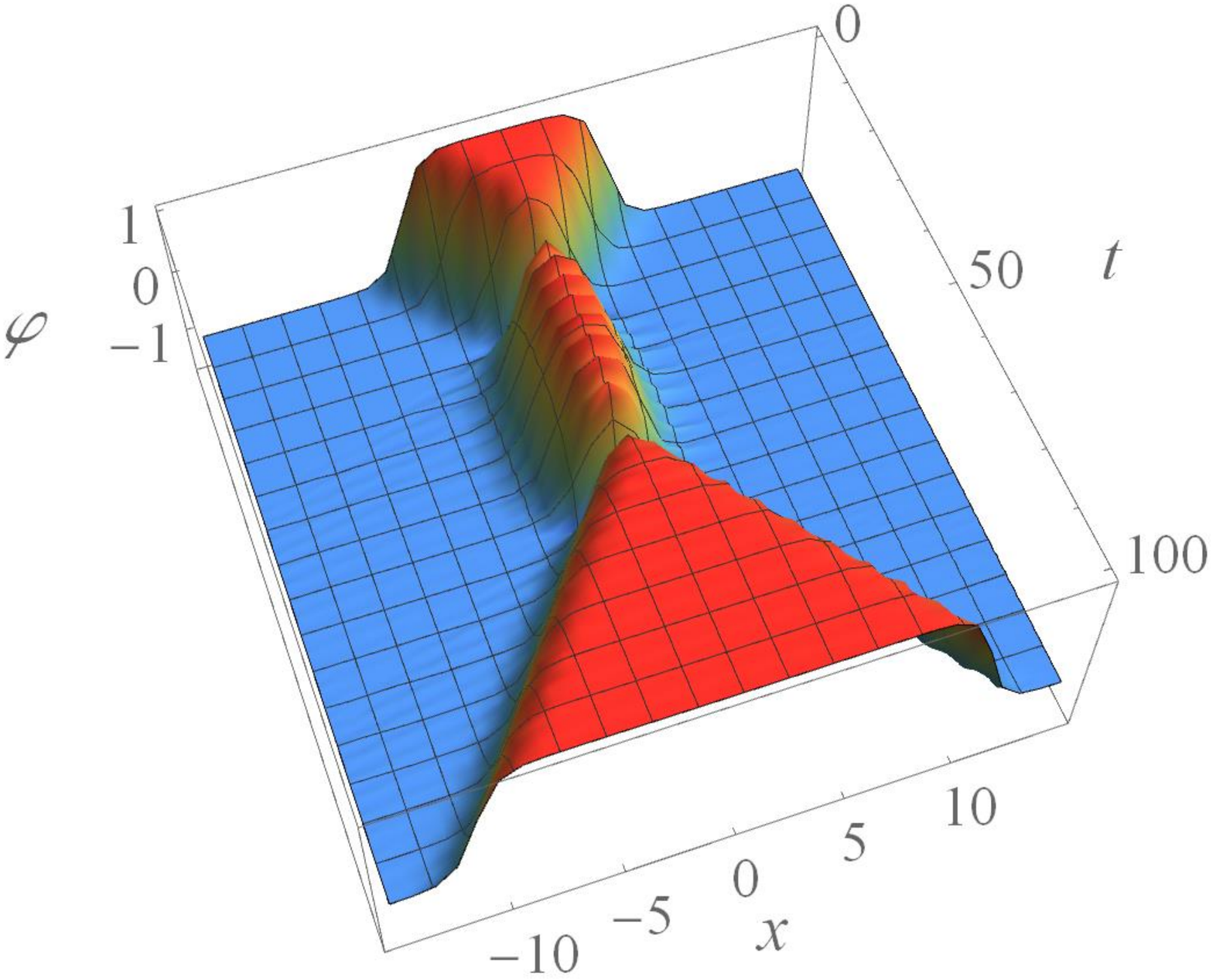}
\end{minipage}
\hfill
\hspace{5mm}
\begin{minipage}{0.47\linewidth}
\centering\includegraphics[width=\linewidth]{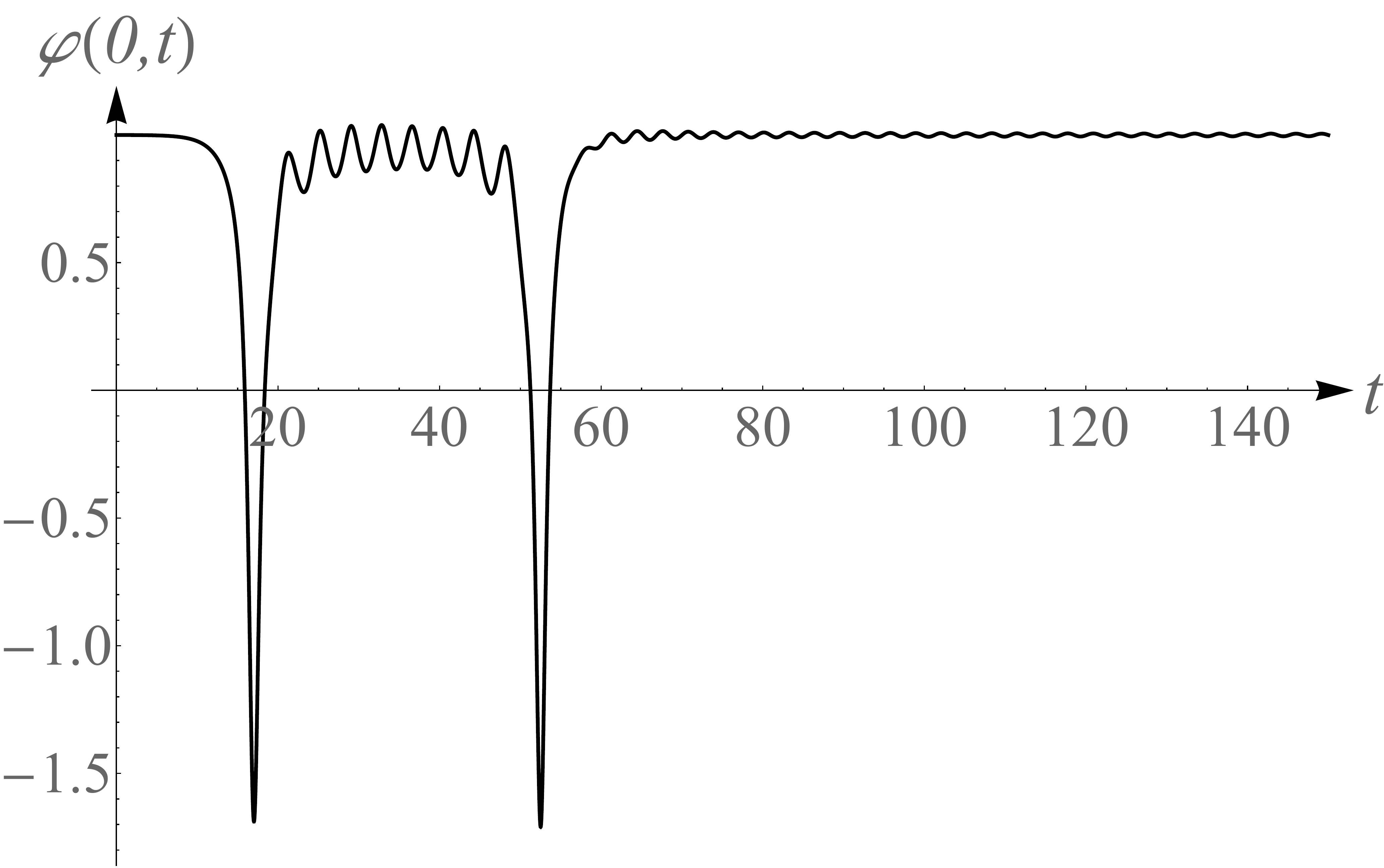}
\end{minipage}
\label{fig:two_bounce_window_phi-4}
}
\vfill
\subfigure[\ Three-bounce window, $v_\mathrm{in}^{}=0.1916$]{
\begin{minipage}{0.47\linewidth}
\centering\includegraphics[width=\linewidth]{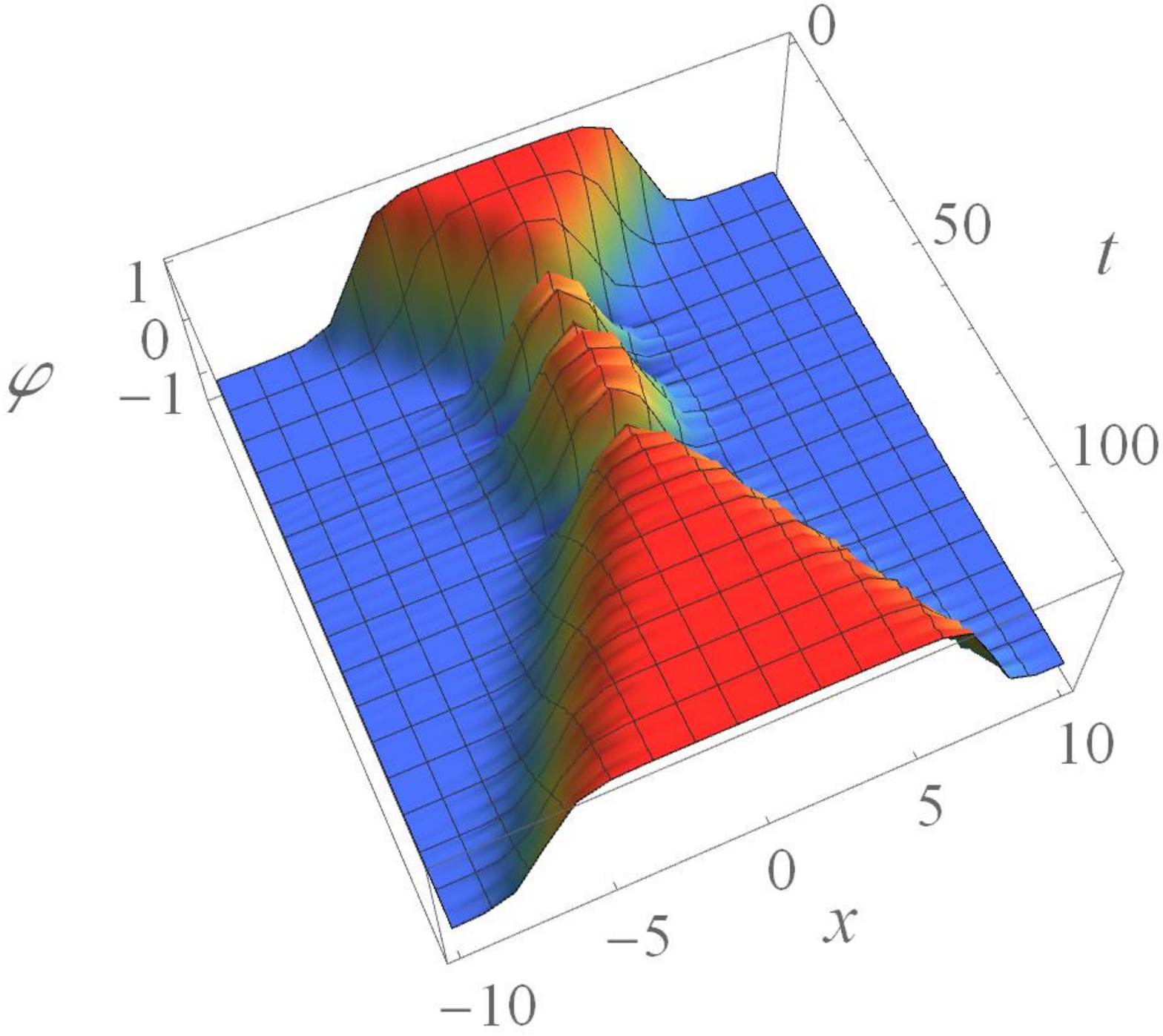}
\end{minipage}
\hfill
\hspace{5mm}
\begin{minipage}{0.47\linewidth}
\centering\includegraphics[width=\linewidth]{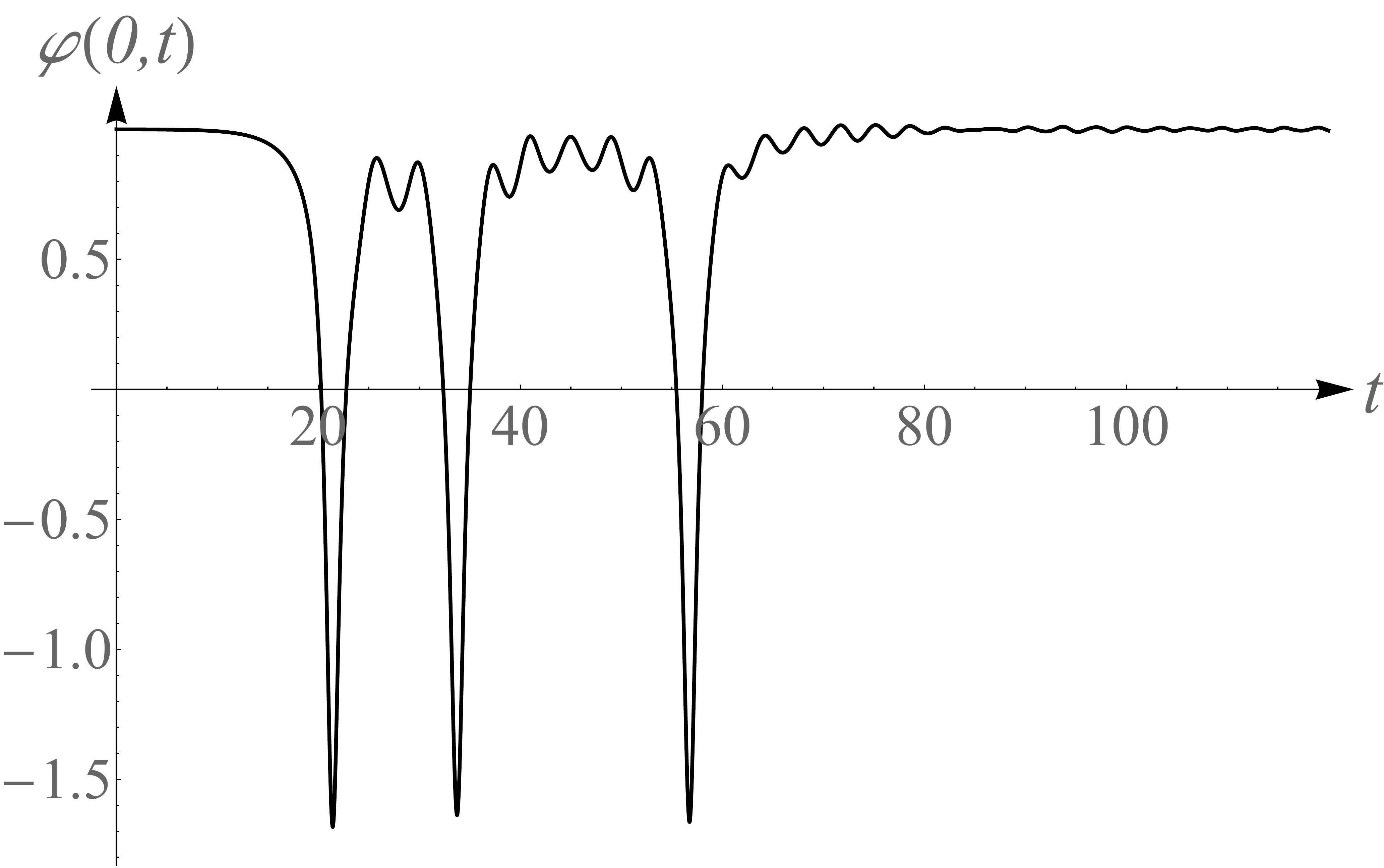}
\end{minipage}
\label{fig:three_bounce_window_phi-4}
}
\vfill
\subfigure[\ Four-bounce window, $v_\mathrm{in}^{}=0.2504$]{
\begin{minipage}{0.47\linewidth}
\centering\includegraphics[width=\linewidth]{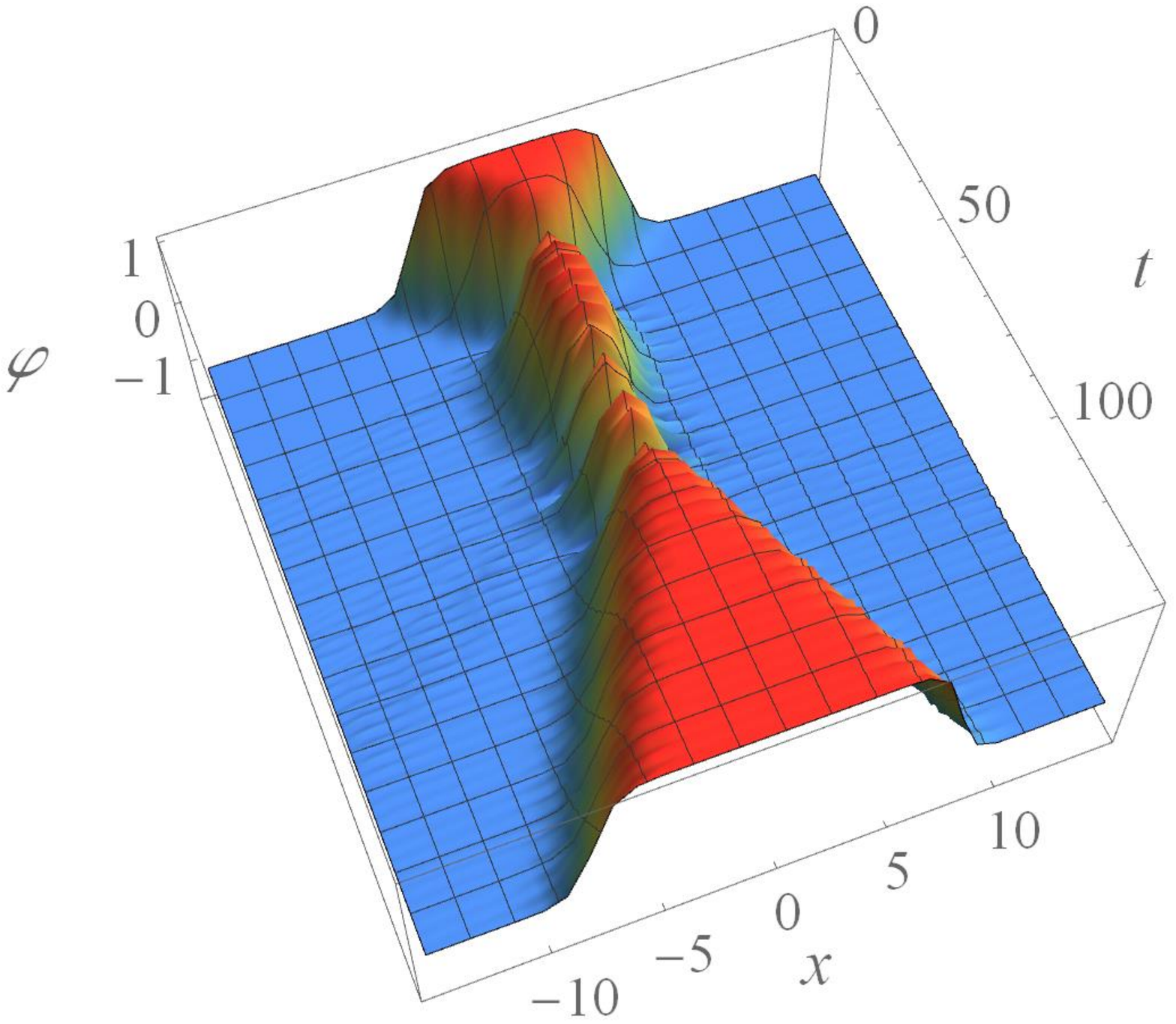}
\end{minipage}
\hfill
\hspace{5mm}
\begin{minipage}{0.47\linewidth}
\centering\includegraphics[width=\linewidth]{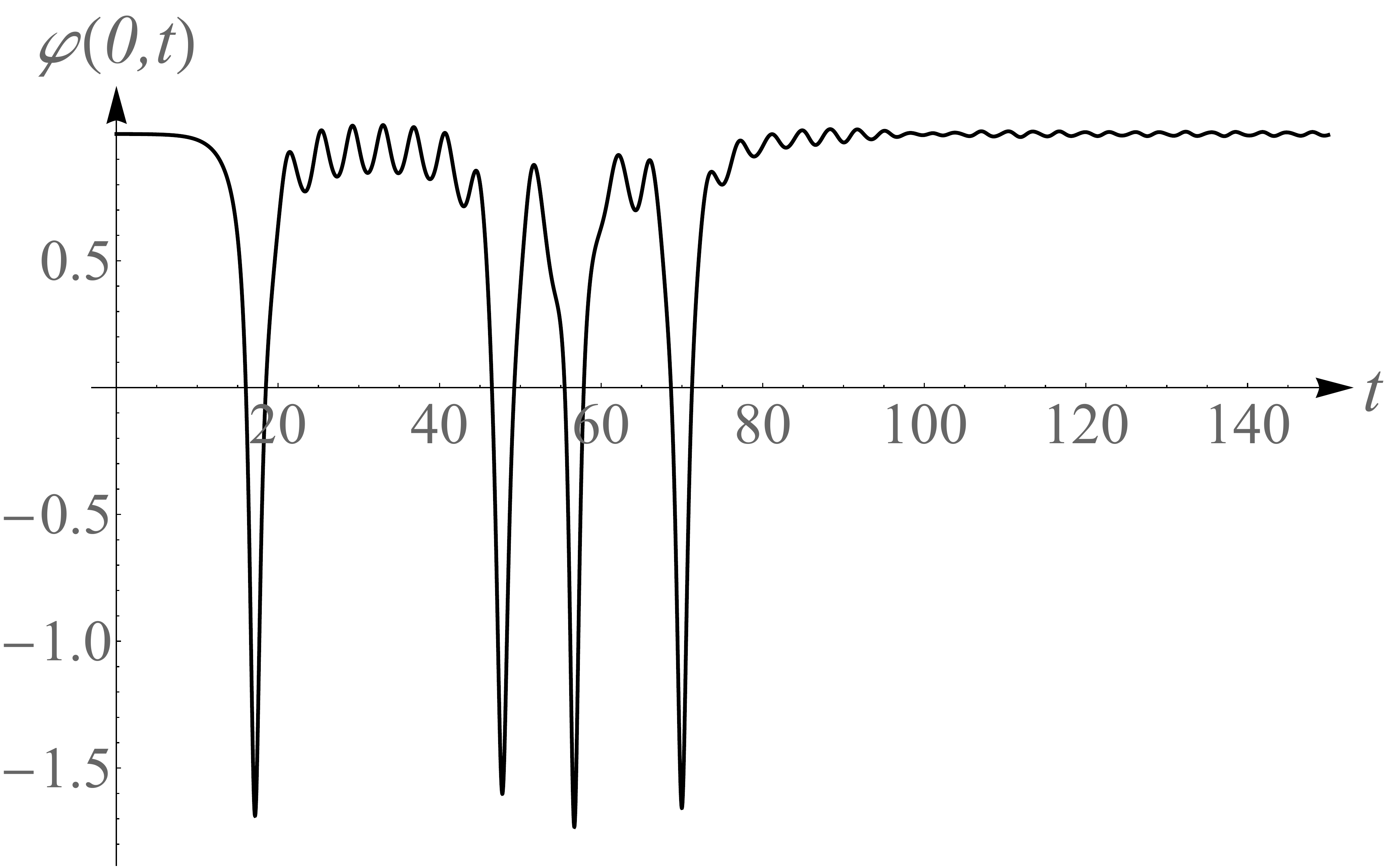}
\end{minipage}
\label{fig:four_bounce_window_phi-4}
}
\caption{Examples of two-, three-, and four-bounce windows in the collisions of the $\varphi^4$ kinks. {\bf Left panel} --- the space-time picture of the escape windows. {\bf Right panel} --- the time dependence of the field at the origin.}
\label{fig:escape_windows_phi-4}
\end{figure}

At $v_\mathrm{in}^{}<v_\mathrm{cr}^{}$ the kinks collide and form a long-living bound state, a bion, which is illustrated in figure \ref{fig:bion_phi-4}.
This bion decays slowly, emitting its energy in the form of waves of small amplitude. However the kinks capture appears not for all $v_\mathrm{in}^{}<v_\mathrm{cr}^{}$, since there is a pattern of {\it escape windows} in the collision processes. An escape window refers to a narrow interval of initial velocities, within which kinks do not form a bound state but escape to infinities. It is important point that, unlike bouncing off at $v_\mathrm{in}^{}>v_\mathrm{cr}^{}$, within an escape window the kinks escape to infinities after two, three or more collisions. According to the number of collisions before escaping, there are two-bounce windows, three-bounce windows, and so on. See figure \ref{fig:escape_windows_phi-4} for some illustrations of two-, three- and four-bounce windows.
The escape windows form a fractal structure. Two-bounce windows are the broadest, and near each of them there is a series of three-bounce windows. Near each three-bounce window, in turn, there is a series of four-bounce windows, and so on, see, e.g., \cite{Kudryavtsev.UFN.1997.eng,Kudryavtsev.UFN.1997.rus}.

The explanation of the appearance of the escape windows is that they are related to the resonance energy exchange between the kinetic energy (the translational mode) and the vibrational mode of the kink (antikink). The mechanism works as follows: consider, for example, the two-bounce window illustrated in figure \ref{fig:two_bounce_window_phi-4}. At the first collision, some part of the kinks kinetic energy is transferred into their vibrational modes. As a result of the loss of the kinetic energy, the kink and the antikink are not able to overcome mutual attraction, and they return and collide again. However, if a certain resonance relation between the time $T_{12}^{}$  between the first and the second collisions and the frequency $\omega_1^{}$ of the vibrational mode holds, a part of the energy can be returned into the kinetic energy, and the kinks are then able to escape from each other.


\section{Deformation procedure and the sinh-deformed $\varphi^4$ model}
\label{sec:sinh-phi-4}

The sinh-deformed $\varphi^4$ model can be obtained from the $\varphi^4$ model by applying the deformation procedure used in refs.~\cite{Bazeia.PRD.2002,Bazeia.PRD.2004,Bazeia.PRD.2006.braneworlds,Bazeia.PRD.2006,Bazeia.IJMPA.2017}. The potential $U_2^{}(\varphi)$ of the new model is related with the old model potential $U_1^{}(\varphi)$ by a deforming function $f(\varphi)$,
\begin{equation}\label{deform}
U_2^{}(\varphi) = \frac{U_1^{}(\varphi\to f(\varphi))}{(df/d\varphi)^2},
\end{equation}
where ``$\varphi\to f(\varphi)$'' means that one must substitute the field $\varphi$ by $f(\varphi)$. At the same time, the kink of the new model, $\varphi_\mathrm{k}^{\mathrm{(new)}}(x)$, can be easily obtained from the kink of the old model, $\varphi_\mathrm{k}^{\mathrm{(old)}}(x)$, by the inverse deforming function $f^{-1}$,
\begin{equation}
\varphi_\mathrm{k}^{\mathrm{(new)}}(x) = f^{-1}(\varphi_\mathrm{k}^{\mathrm{(old)}}(x)).
\end{equation}

We start from the $\varphi^4$ model with the potential \eqref{eq:potential_phi-4} and use the deforming function $f(\varphi)=\sinh\varphi$. Then we come to the potential of the sinh-deformed $\varphi^4$ model
\begin{equation}
\label{eq:potential_sinh_phi4}
U_2^{}(\varphi) = \frac{1}{2}\:\mbox{sech}^2\varphi\left(1-\sinh^2\varphi\right)^2.
\end{equation}
This potential has two degenerate minima, $\varphi_\pm^{}=\pm\mbox{arsinh}\:1$, $U_2(\varphi_\pm^{})=0$, see figure \ref{fig:potentials}. The kinks of the sinh-deformed $\varphi^4$ model are
\begin{equation}
\label{eq:sinh_phi4_kinks}
\varphi_\mathrm{k}^{}(x) = \mbox{arsinh}(\tanh x), \quad \varphi_\mathrm{\bar k}^{}(x) = -\mbox{arsinh}(\tanh x),
\end{equation}
see figure \ref{fig:kinks}. The mass of the sinh-deformed $\varphi^4$ kink (antikink), i.e.\ the energy $E[\varphi_\mathrm{k}^{}(x)]$ of the static kink (antikink), is
\begin{equation}
M_\mathrm{k} = \pi-2.
\end{equation}

\begin{figure}[h!]
\centering
\includegraphics[scale=0.2]{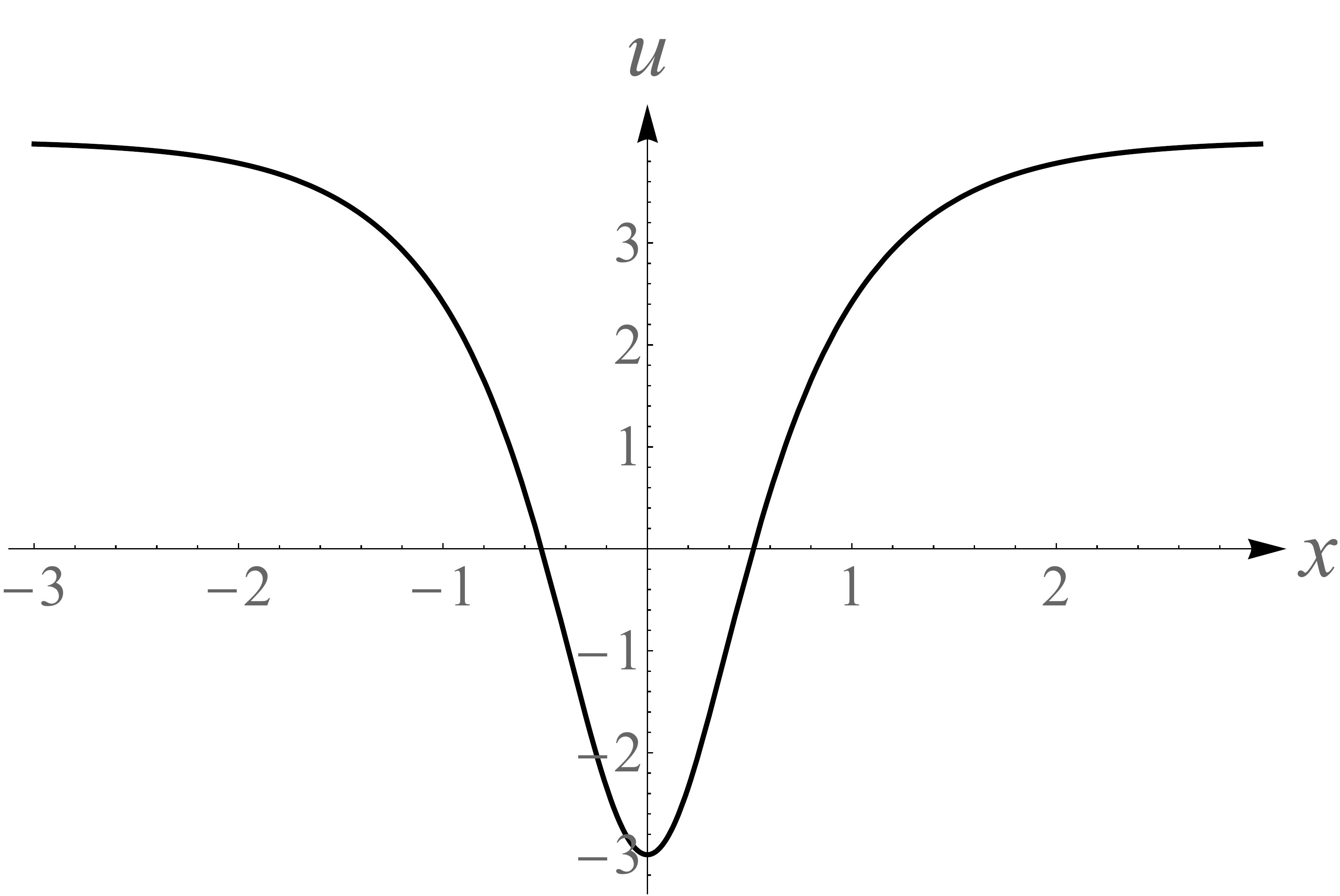}
\caption{The quantum-mechanical potential \eqref{eq:Schrod_potential_sinh_phi4}.}
\label{fig:Schrod_potential_sinh_phi4}
\end{figure}

The excitation spectrum of the kink (antikink) \eqref{eq:sinh_phi4_kinks} is defined by the quantum-mechanical potential
\begin{equation}\label{eq:Schrod_potential_sinh_phi4}
u_2^{}(x) = 2\tanh^2x+1+\frac{8\tanh^2x-4}{(1+\tanh^2x)^2},
\end{equation}
which is presented in figure \ref{fig:Schrod_potential_sinh_phi4}.
We performed a numerical search of the discrete part of the excitation spectrum in the potential \eqref{eq:Schrod_potential_sinh_phi4}. This problem was solved using the standard {\it shooting} method. For various values of $\omega^2$ we integrated eq.~\eqref{eq:Schrodinger} with the known asymptotic behavior $\eta(x)\sim\exp(-\sqrt{4-\omega^2}|x|)$ of its solutions at $x\to\pm\infty$, starting from a large negative $x$ and from a large positive $x$. As a result, we obtained two different solutions, the ``left'' solution and the ``right'' solution, which were then matched at some point $\bar{x}$ close to the origin (the particular choice of $\bar{x}$ is not important). The Wronskian of the ``left'' and the ``right'' solutions, calculated at the matching point, as a function of $\omega^2$ turns to zero at eigenvalues of the Hamiltonian \eqref{eq:Schrod_Ham} with the potential \eqref{eq:Schrod_potential_sinh_phi4}.

We found two levels in the potential \eqref{eq:Schrod_potential_sinh_phi4}: the zero mode $\omega_0^{}=0$, and the vibrational mode with the frequency $\omega_1^{}\approx 1.89$.

\section{Kink-antikink collisions in the sinh-deformed $\varphi^4$ model}
\label{sec:Scattering_sinh_phi-4}

We studied the collisions of the kink and the antikink of the sinh-deformed $\varphi^4$ model using the initial configuration similar to that used in section \ref{sec:phi-4} in the case of the $\varphi^4$ kinks scattering, namely
\begin{equation}\label{eq:in_cond_sinh_phi-4}
\varphi(x,t) = \mbox{arsinh}\left[\tanh\left(\frac{x+x_0^{}-v_\mathrm{in}^{}t }{\sqrt{1-v_\mathrm{in}^2}}\right)\right] - \mbox{arsinh}\left[\tanh\left(\frac{x-x_0^{}+v_\mathrm{in}^{}t}{\sqrt{1-v_\mathrm{in}^2}}\right)\right] - \mbox{arsinh}\:1,
\end{equation}
which corresponds to the kink and the antikink centered at $x=\pm x_0^{}$ and moving towards each other with the initial velocities $v_\mathrm{in}^{}$. We used $2x_0^{}=10$ and the same parameters of the numerical scheme as for the $\varphi^4$ kinks in section \ref{sec:phi-4}, see eq.~\eqref{eq:difference_scheme} and the paragraph below this equation.

\begin{figure}[h!]
\begin{minipage}{0.47\linewidth}
\centering\includegraphics[width=\linewidth]{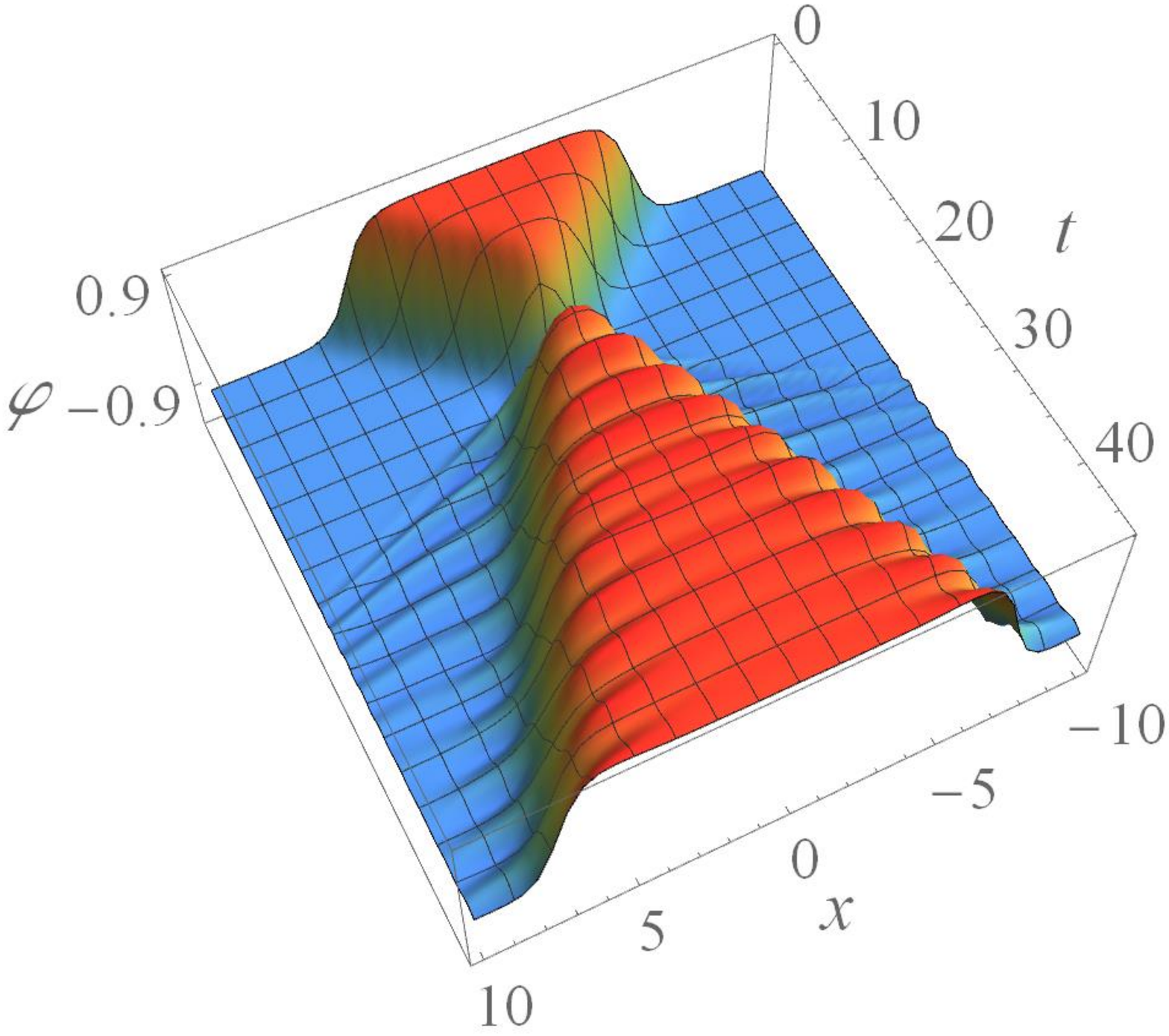}
\end{minipage}
\hspace{5mm}
\begin{minipage}{0.47\linewidth}
\centering\includegraphics[width=\linewidth]{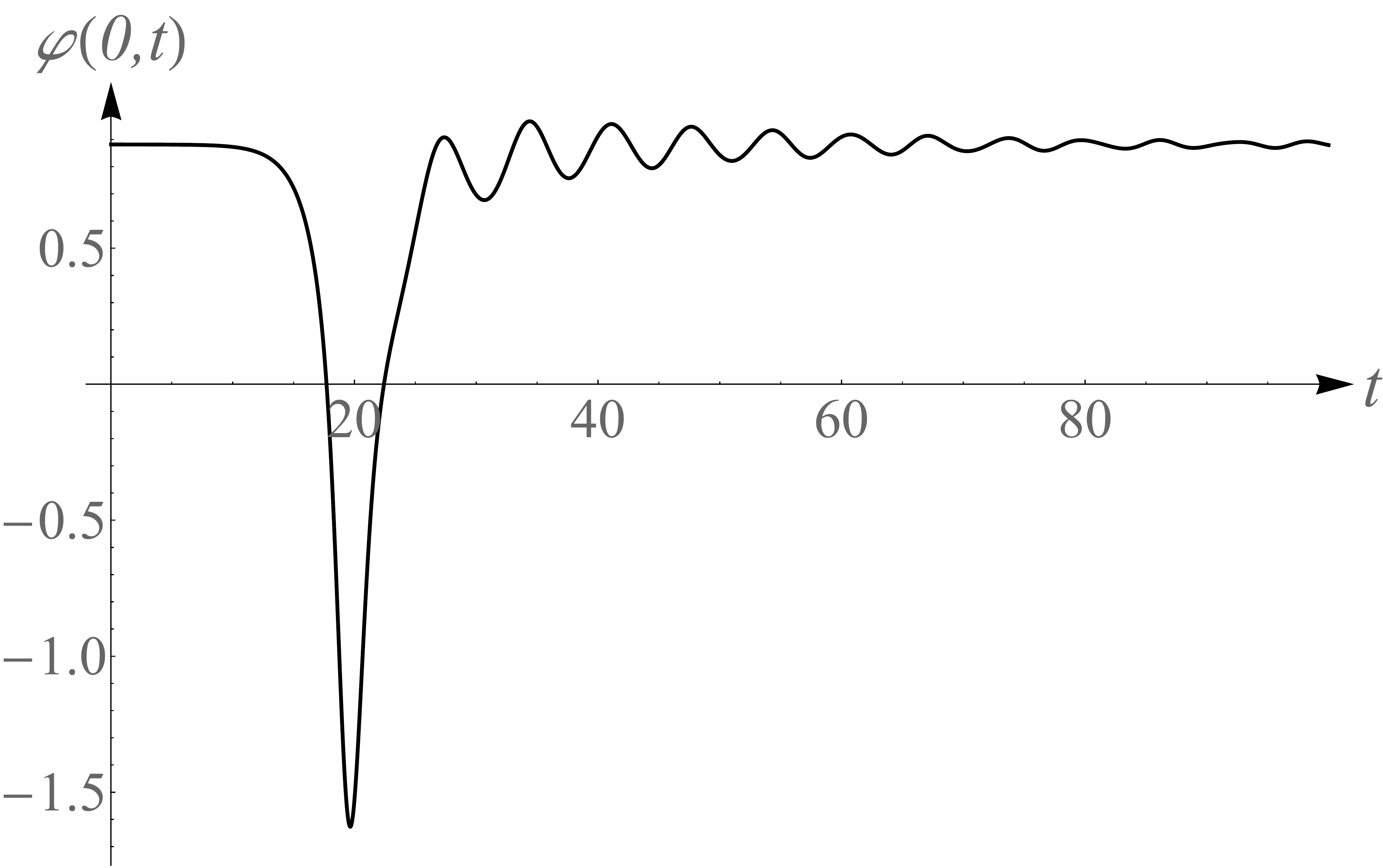}
\end{minipage}
\caption{Bounce off and escape of kinks of the sinh-deformed $\varphi^4$ model at $v_\mathrm{in}=0.5000$. {\bf Left panel} --- the space-time picture of the field evolution. {\bf Right panel} --- the time dependence of the field at the origin.}
\label{fig:escape_sinh_phi-4}
\end{figure}

We found a critical value of the initial velocity $v_\mathrm{cr}^{}\approx 0.4639$, which separates two different regimes of the kinks scattering. At $v_\mathrm{in}^{}>v_\mathrm{cr}^{}$ the kinks bounce off and escape to infinities after one collision. This is illustrated in figure \ref{fig:escape_sinh_phi-4}, and the situation here is similar to that observed for the $\varphi^4$ kinks above the critical velocity, as depicted in figure \ref{fig:escape_phi-4}.

\begin{figure}[h!]
\begin{minipage}{0.47\linewidth}
\centering\includegraphics[width=\linewidth]{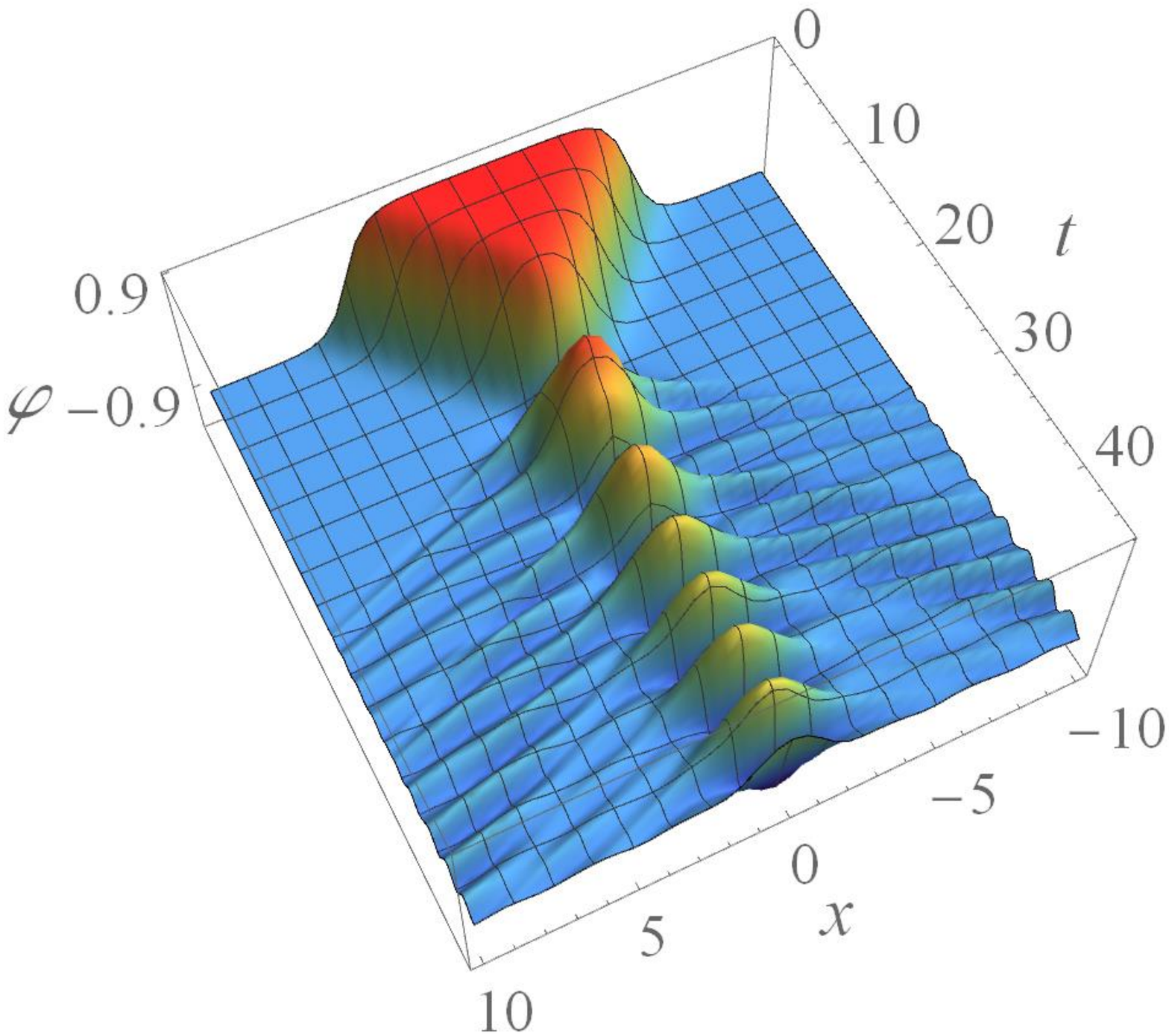}
\end{minipage}
\hspace{5mm}
\begin{minipage}{0.47\linewidth}
\centering\includegraphics[width=\linewidth]{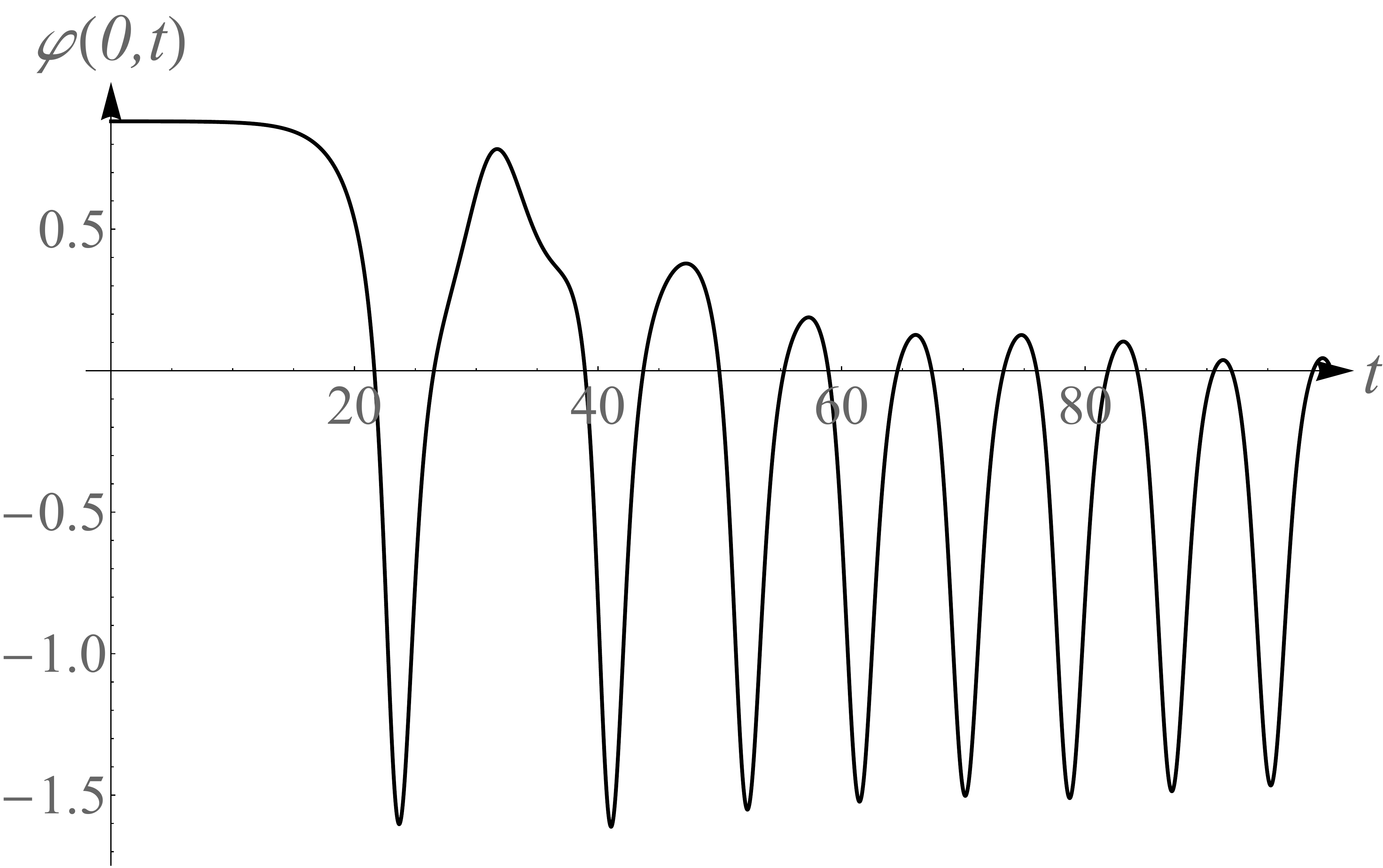}
\end{minipage}
\caption{The formation of a bound state in the sinh-deformed $\varphi^4$ kink-antikink collision at $v_\mathrm{in}^{}=0.4000$. {\bf Left panel} --- the space-time picture of a bion formation. {\bf Right panel} --- the time dependence of the field at the origin.}
\label{fig:bion_sinh_phi-4}
\end{figure}

\begin{figure}[h!]
\subfigure[\ Two-bounce window, $v_\mathrm{in}^{}=0.4600$]{
\begin{minipage}{0.47\linewidth}
\centering\includegraphics[width=\linewidth]{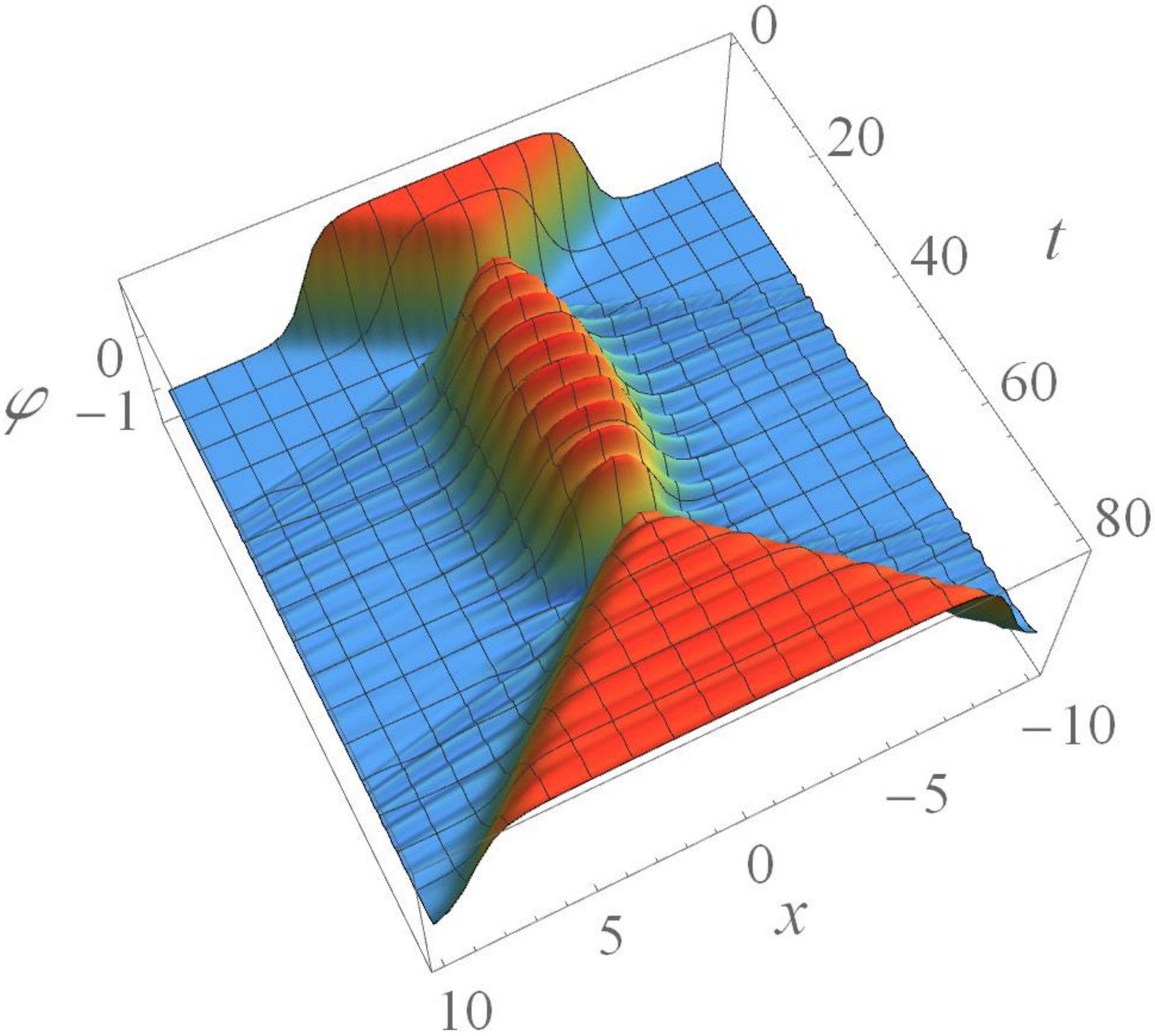}
\end{minipage}
\hspace{5mm}
\begin{minipage}{0.47\linewidth}
\centering\includegraphics[width=\linewidth]{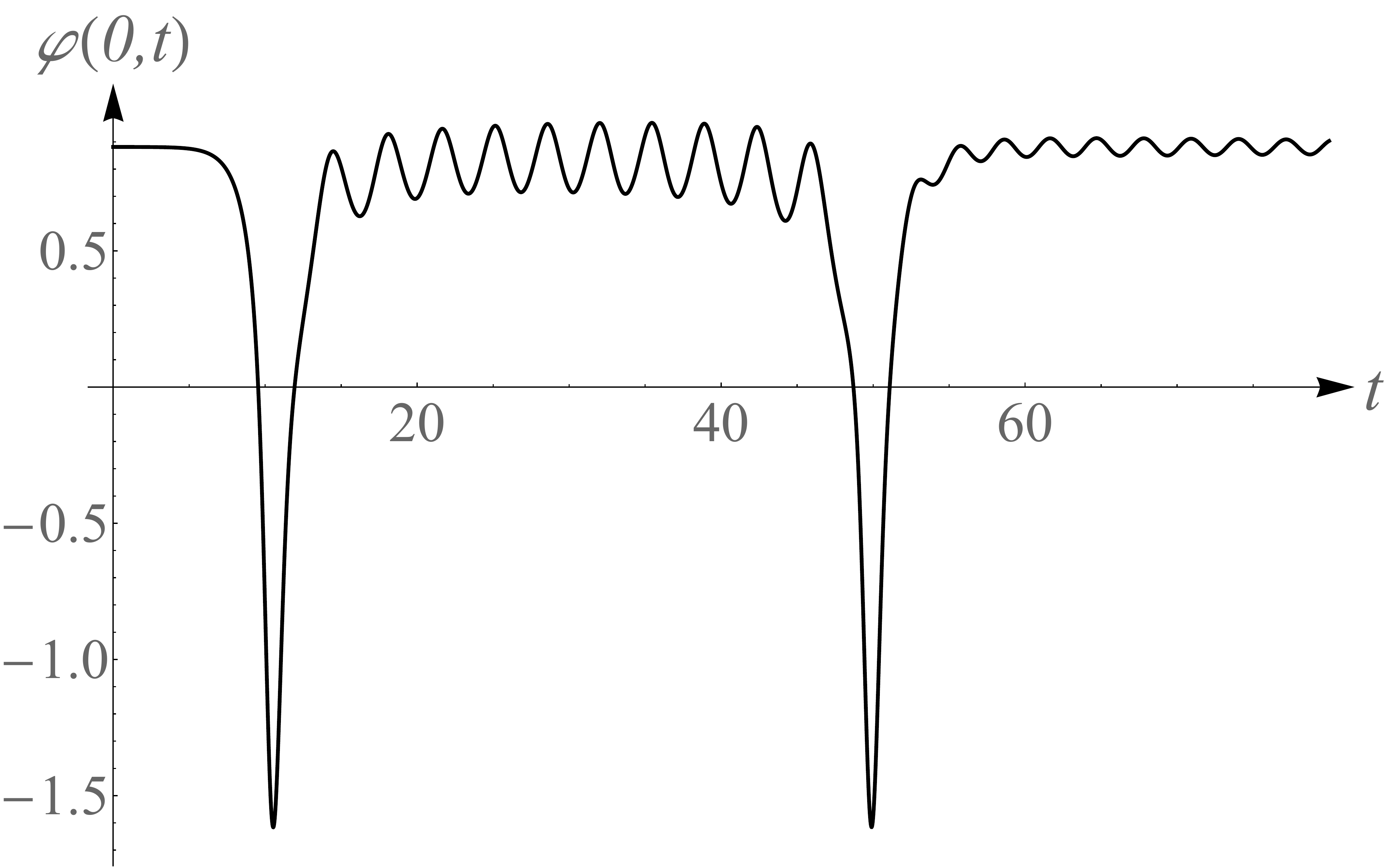}
\end{minipage}
}
\subfigure[\ Three-bounce window, $v_\mathrm{in}^{}=0.4500$]{
\begin{minipage}{0.47\linewidth}
\centering\includegraphics[width=\linewidth]{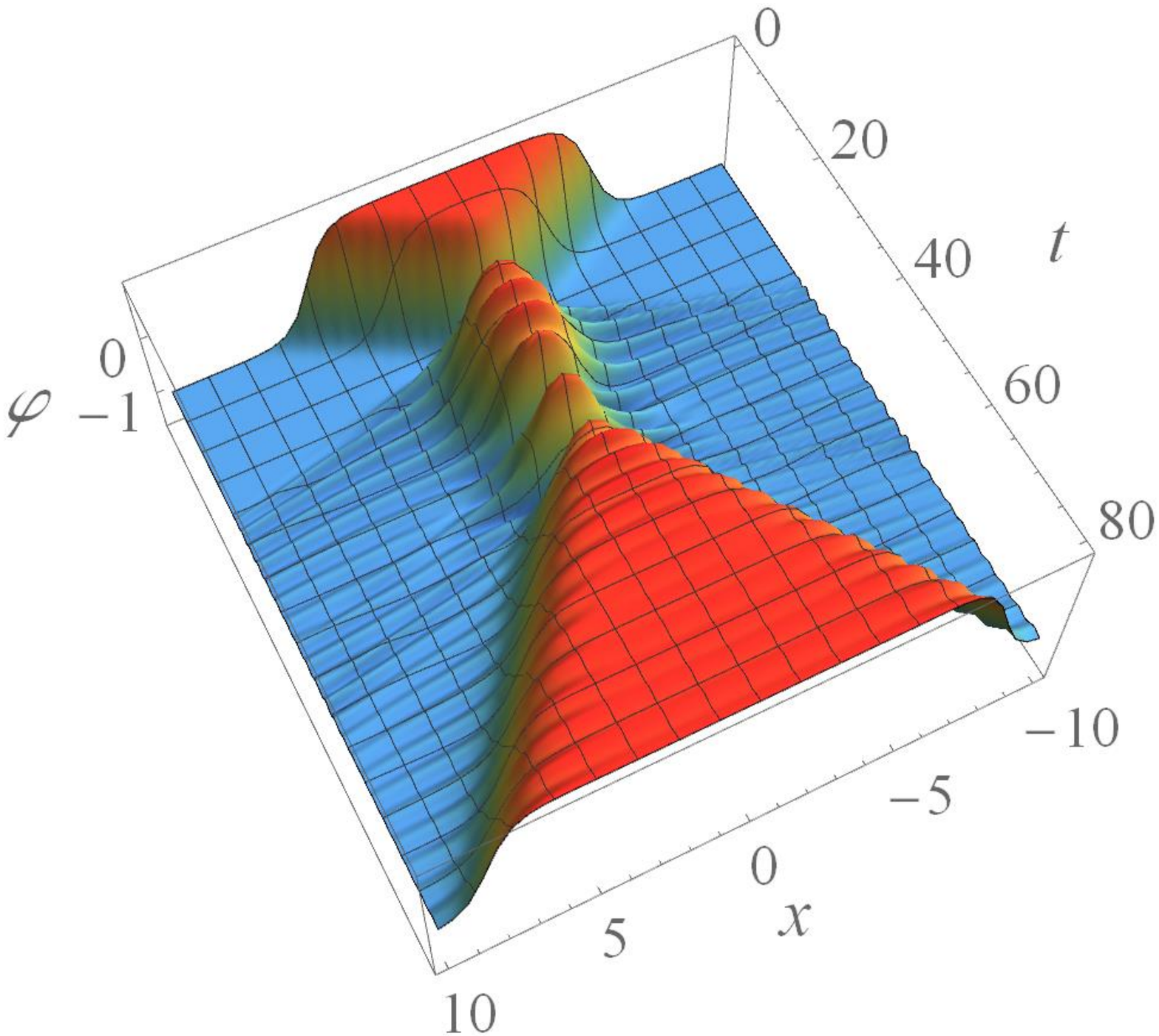}
\end{minipage}
\hspace{5mm}
\begin{minipage}{0.47\linewidth}
\centering\includegraphics[width=\linewidth]{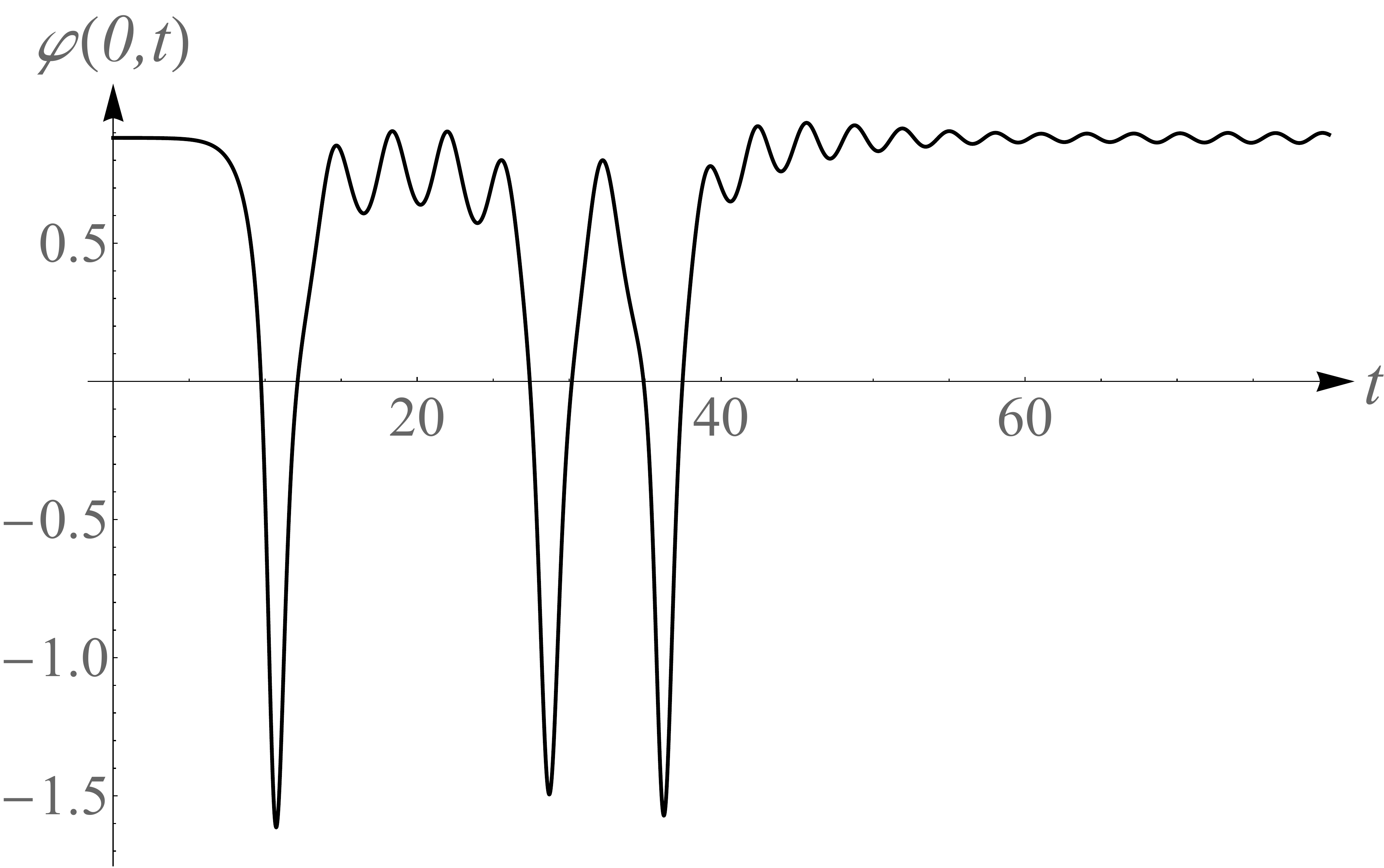}
\end{minipage}
}
\caption{Examples of two- and three-bounce windows in the collisions of kinks of the sinh-deformed $\varphi^4$ model. {\bf Left panel} --- the space-time picture of the escape windows. {\bf Right panel} --- the time dependence of the field at the origin.}
\label{fig:escape_windows_sinh_phi-4}
\end{figure}

\begin{figure}[h!]
\vspace{-25ex}
\centering\includegraphics[width=\linewidth]{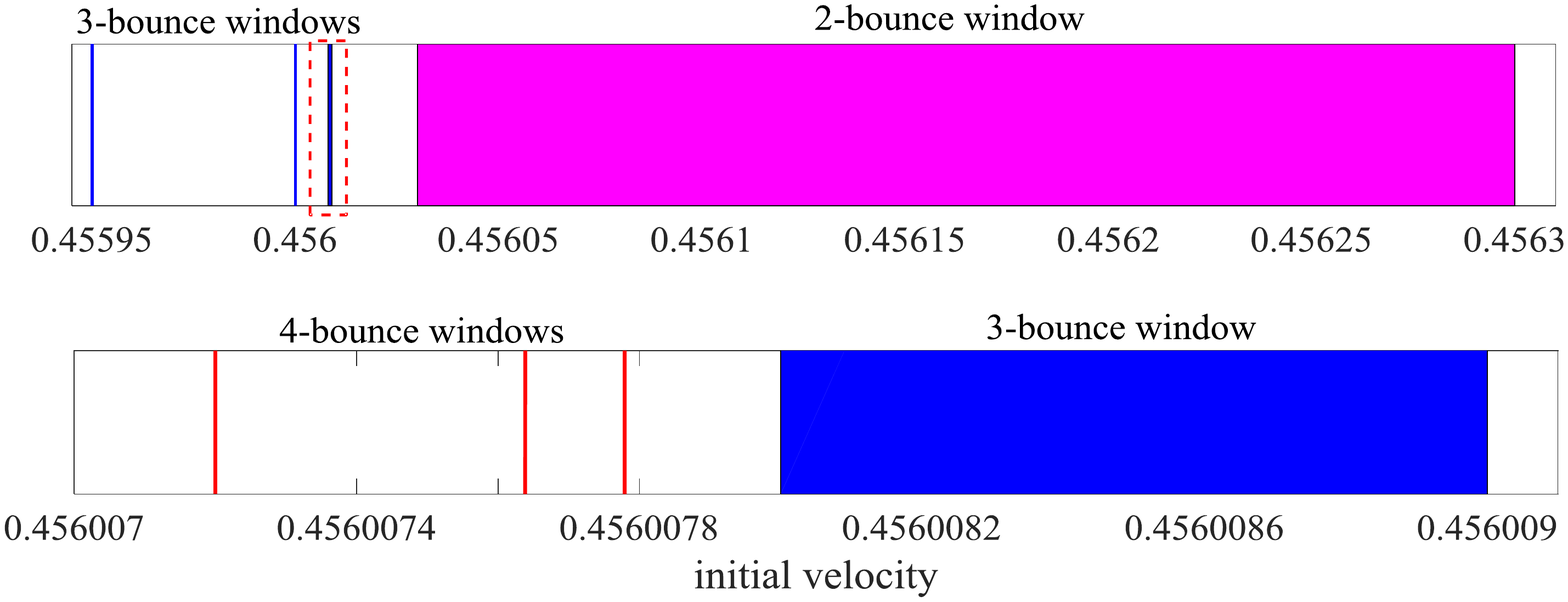}
\caption{{\bf Top panel} --- a two-bounce escape window and some of the nearby three-bounce windows. {\bf Bottom panel} --- a three-bounce window (which is boxed on the top panel) with some nearby four-bounce windows.}
\label{fig:fractal}
\end{figure}

At the initial velocities below the critical value, $v_\mathrm{in}^{}<v_\mathrm{cr}^{}$, we observed the kinks' capture and formation of their bound state, figure \ref{fig:bion_sinh_phi-4},
and a rich variety of resonance phenomena. First of all, in this range of the initial velocities we found a complicated pattern of escape windows, similar to the case of the $\varphi^4$ kinks. We identified many two-bounce, three-bounce, etc., escape windows. The field dynamics within these windows is similar to the case of the $\varphi^4$ model. In figure \ref{fig:escape_windows_sinh_phi-4}
we present examples of the field behavior within two- and three-bounce windows. In a way similar to the case of the $\varphi^4$ model, the escape windows that appear in the sinh-deformed model seem to form a fractal structure. We found several three-bounce escape windows near a two-bounce window, see figure \ref{fig:fractal}.
Near one of the three-bounce escape windows we observed four-bounce escape windows. This behavior is similar to the one found in the $\varphi^4$ model, so it suggests that the escape windows also form a fractal structure in this case.

\begin{figure}[h!]
\subfigure[\ Small amplitude, $v_\mathrm{in}^{}=0.44183$]{
\begin{minipage}{0.47\linewidth}
\centering\includegraphics[width=\linewidth]{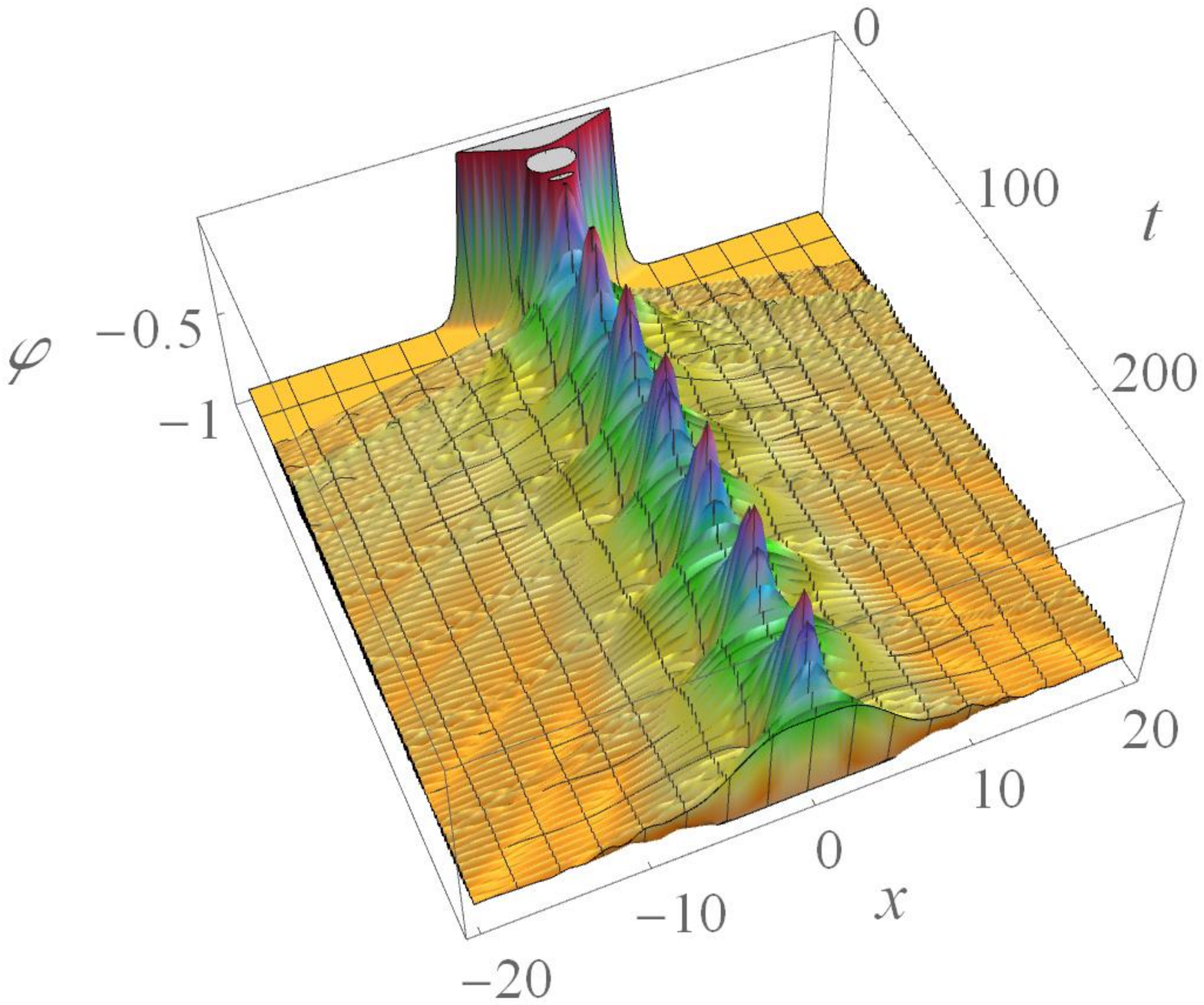}
\end{minipage}
\hspace{5mm}
\begin{minipage}{0.47\linewidth}
\centering\includegraphics[width=\linewidth]{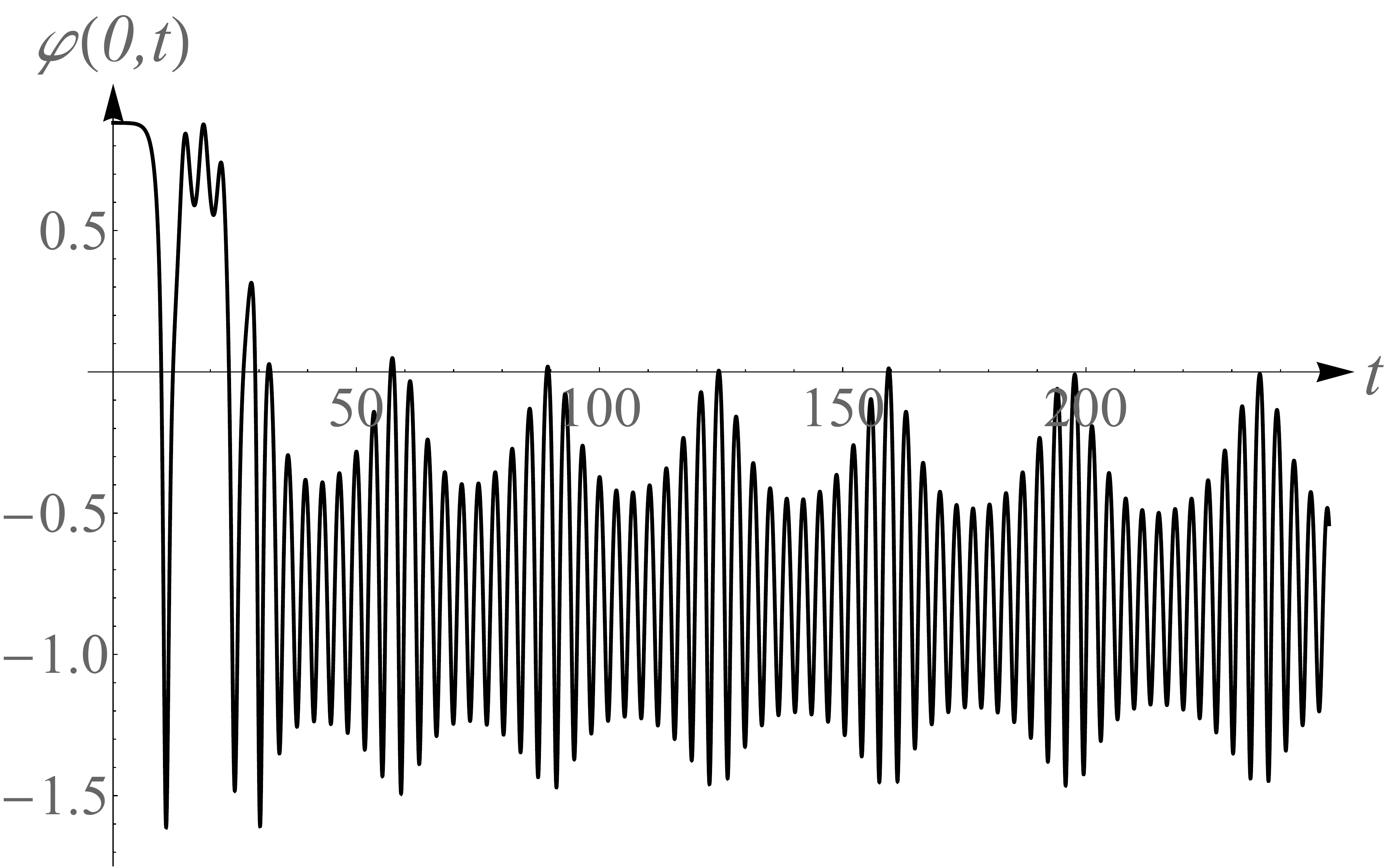}
\end{minipage}
}
\subfigure[\ Medium amplitude, $v_\mathrm{in}^{}=0.44188$]{
\begin{minipage}{0.47\linewidth}
\centering\includegraphics[width=\linewidth]{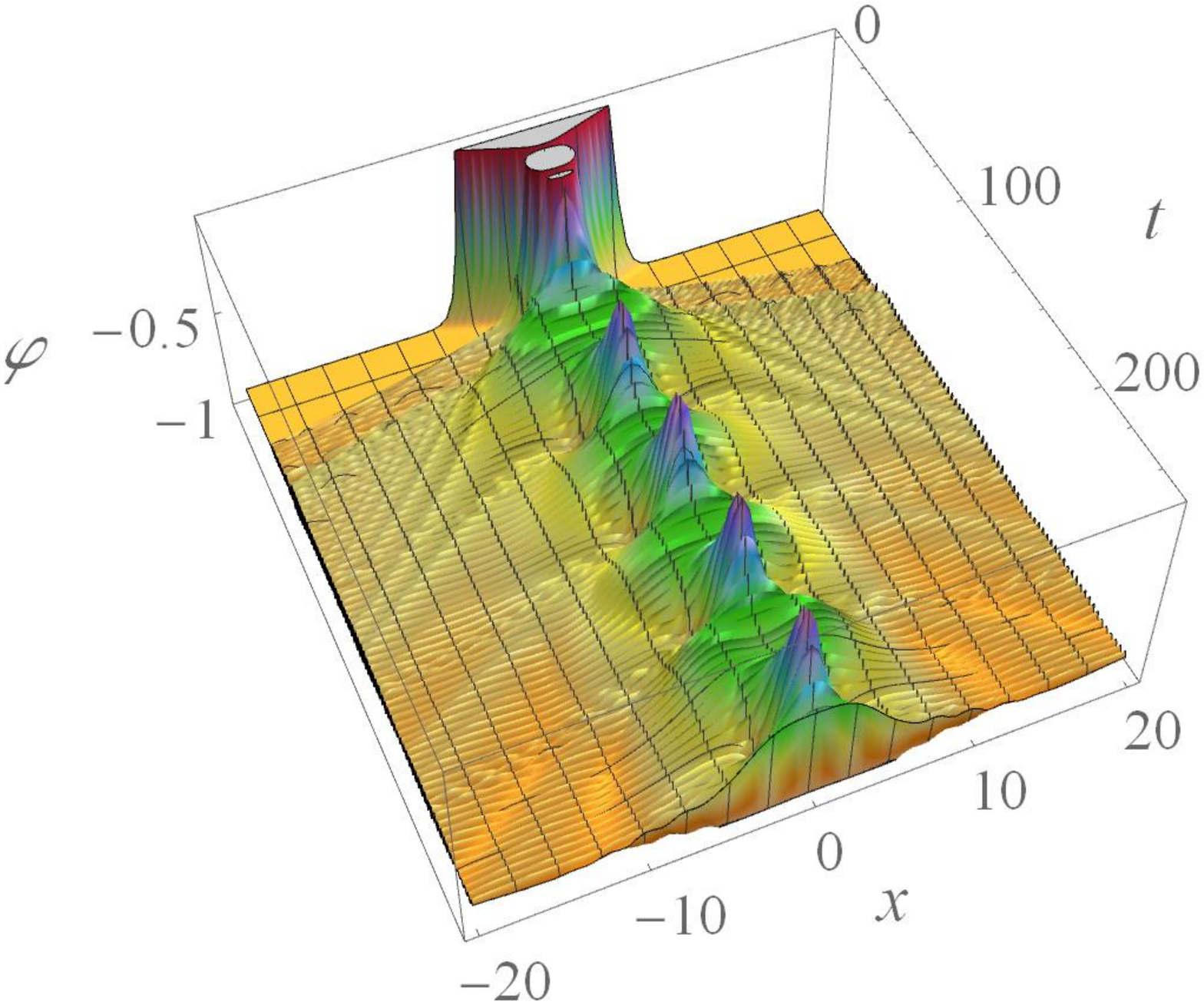}
\end{minipage}
\hspace{5mm}
\begin{minipage}{0.47\linewidth}
\centering\includegraphics[width=\linewidth]{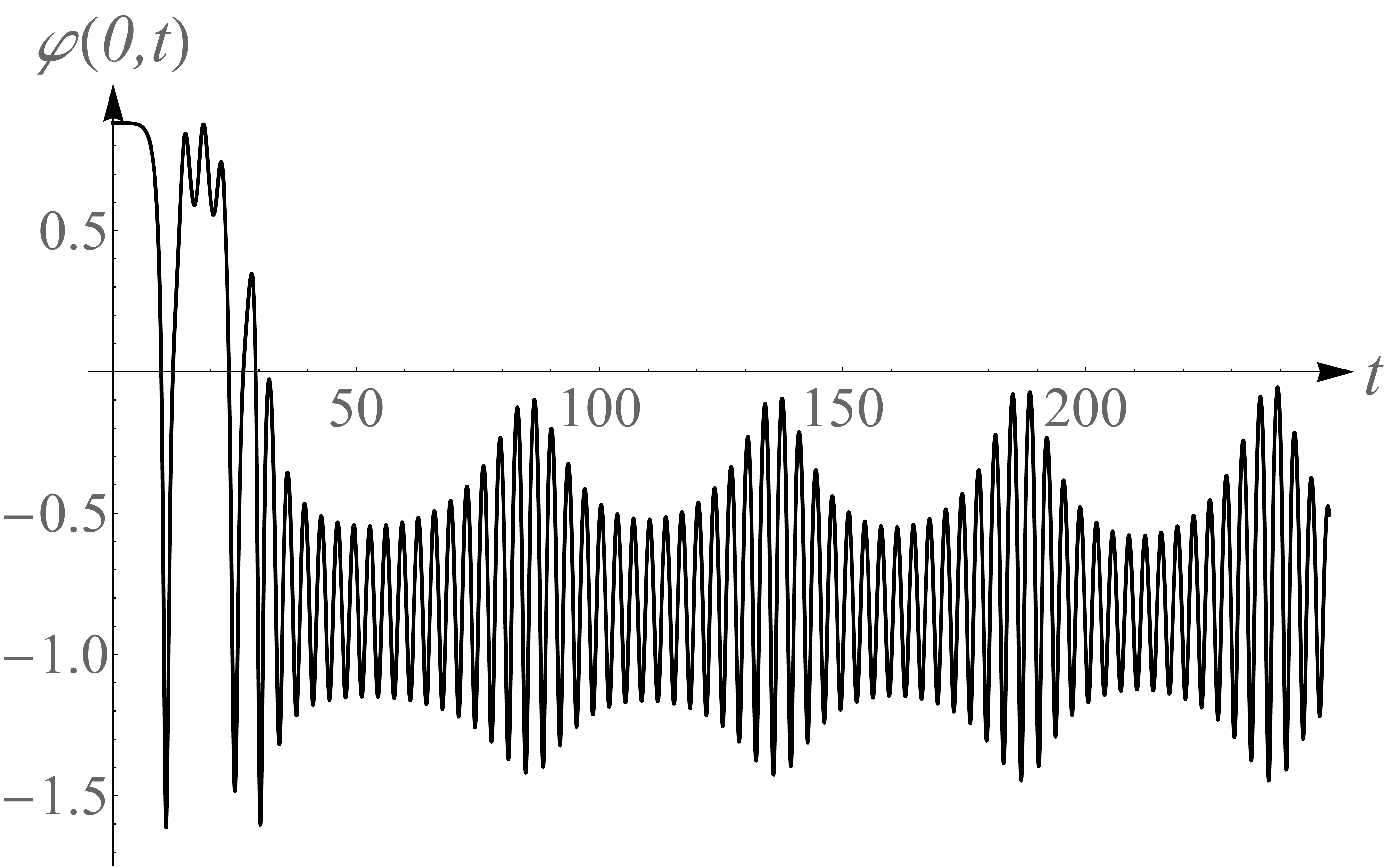}
\end{minipage}
}
\subfigure[\ Large amplitude, $v_\mathrm{in}^{}=0.44190$]{
\begin{minipage}{0.47\linewidth}
\centering\includegraphics[width=\linewidth]{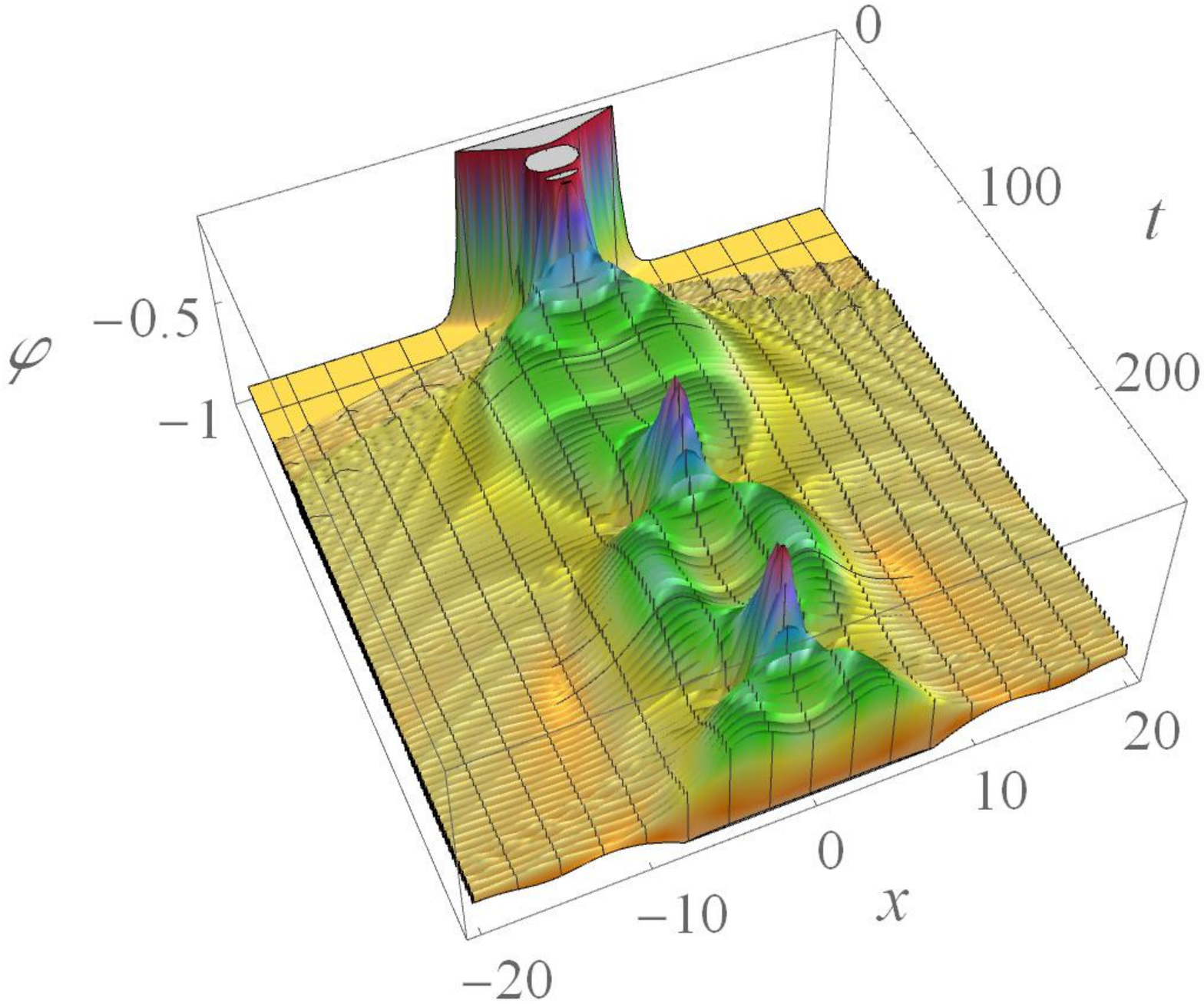}
\end{minipage}
\hspace{5mm}
\begin{minipage}{0.47\linewidth}
\centering\includegraphics[width=\linewidth]{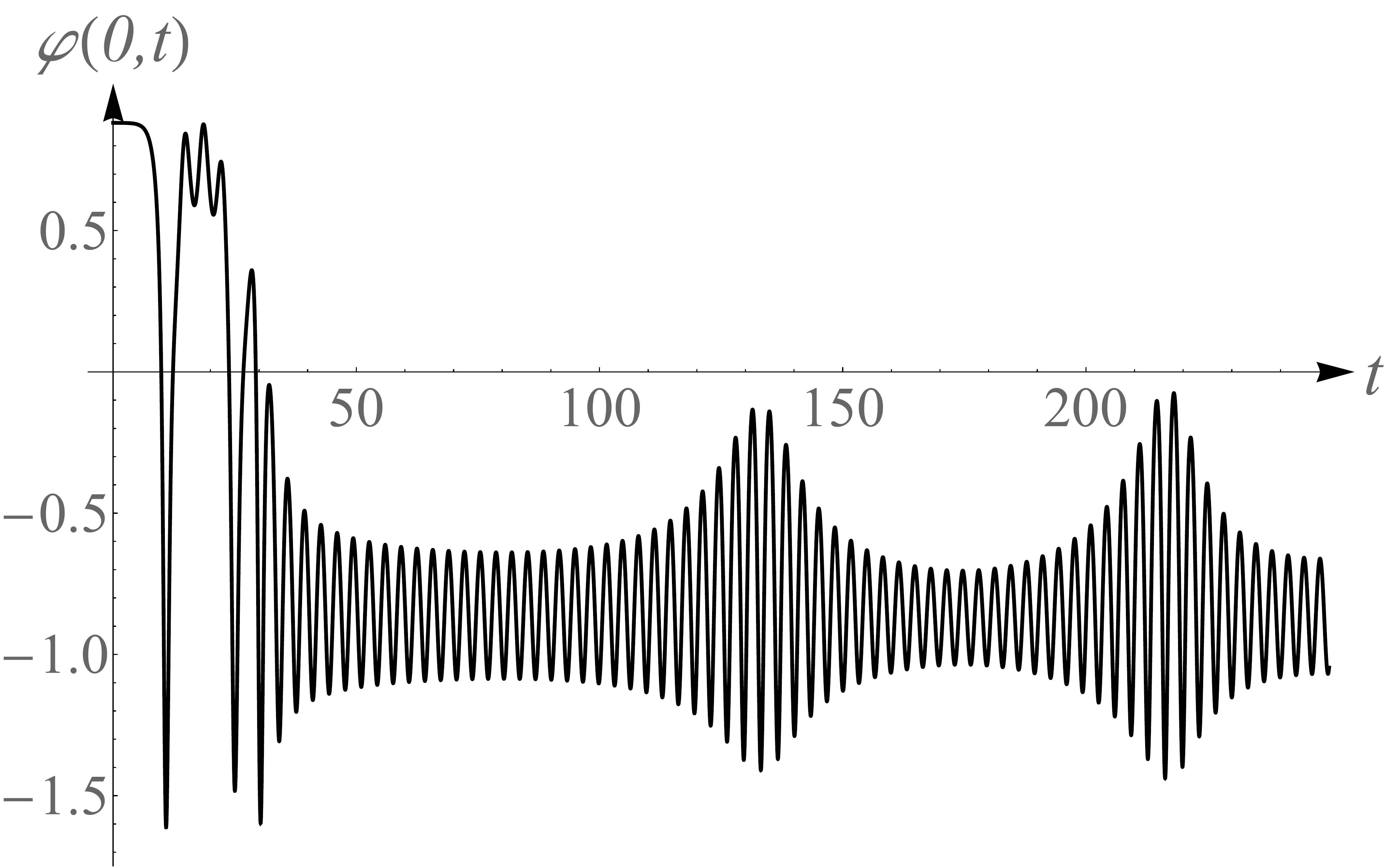}
\end{minipage}
}
\caption{Formation of a bound state of two oscillons in the collisions of kinks of the sinh-deformed $\varphi^4$ model. {\bf Left panel} --- the space-time picture. {\bf Right panel} --- the time dependence of the field at the origin.}
\label{fig:oscillating_bions_sinh_phi-4}
\end{figure}

\begin{figure}[h!]
\subfigure[$v_\mathrm{in}^{}=0.44191$, $v_\mathrm{os}^{}\approx 0.04$]{
\begin{minipage}{0.5\linewidth}
\centering\includegraphics[width=\linewidth]{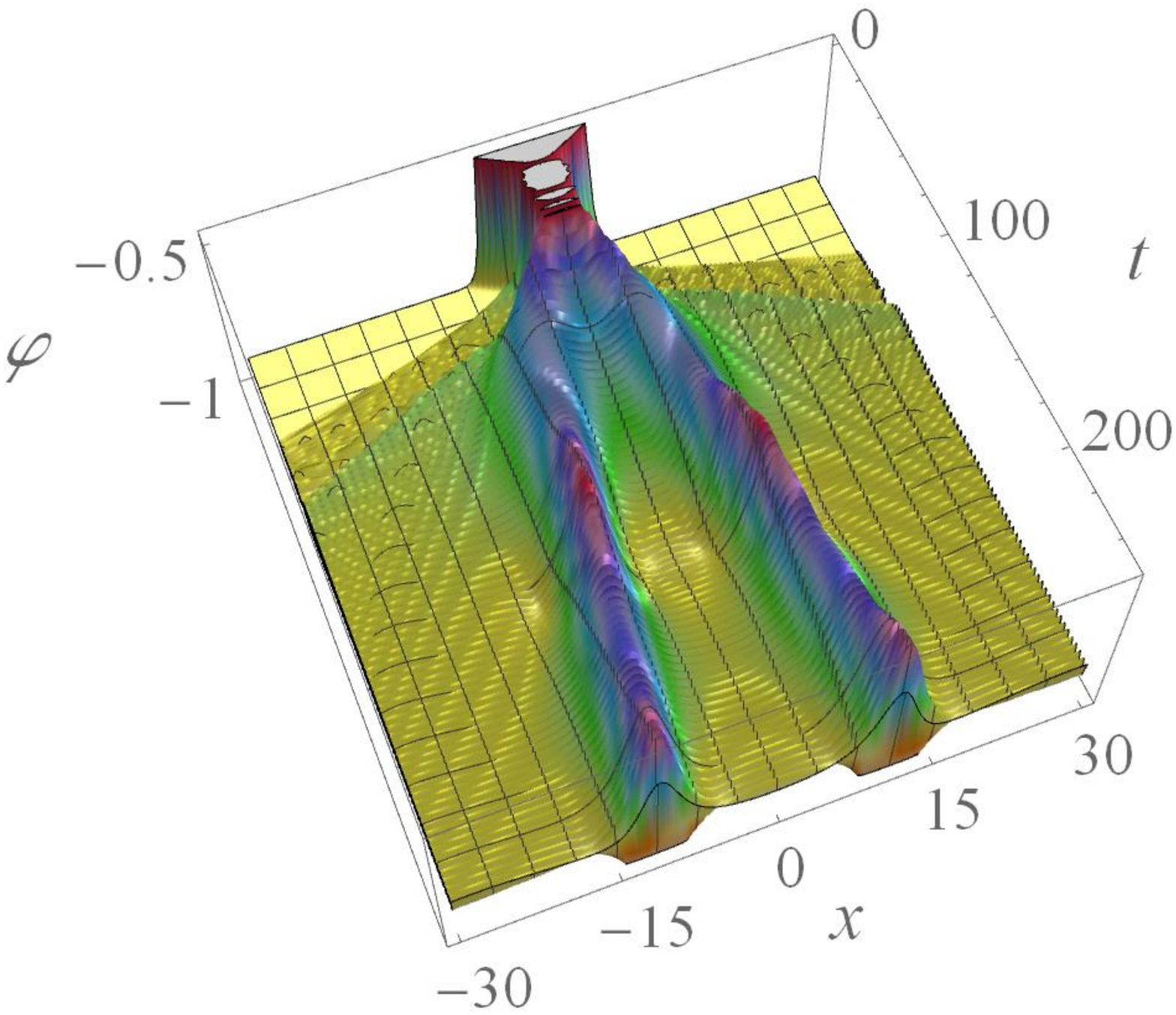}
\end{minipage}
}
\subfigure[$v_\mathrm{in}^{}=0.44195$, $v_\mathrm{os}^{}\approx 0.10$]{
\begin{minipage}{0.5\linewidth}
\centering\includegraphics[width=\linewidth]{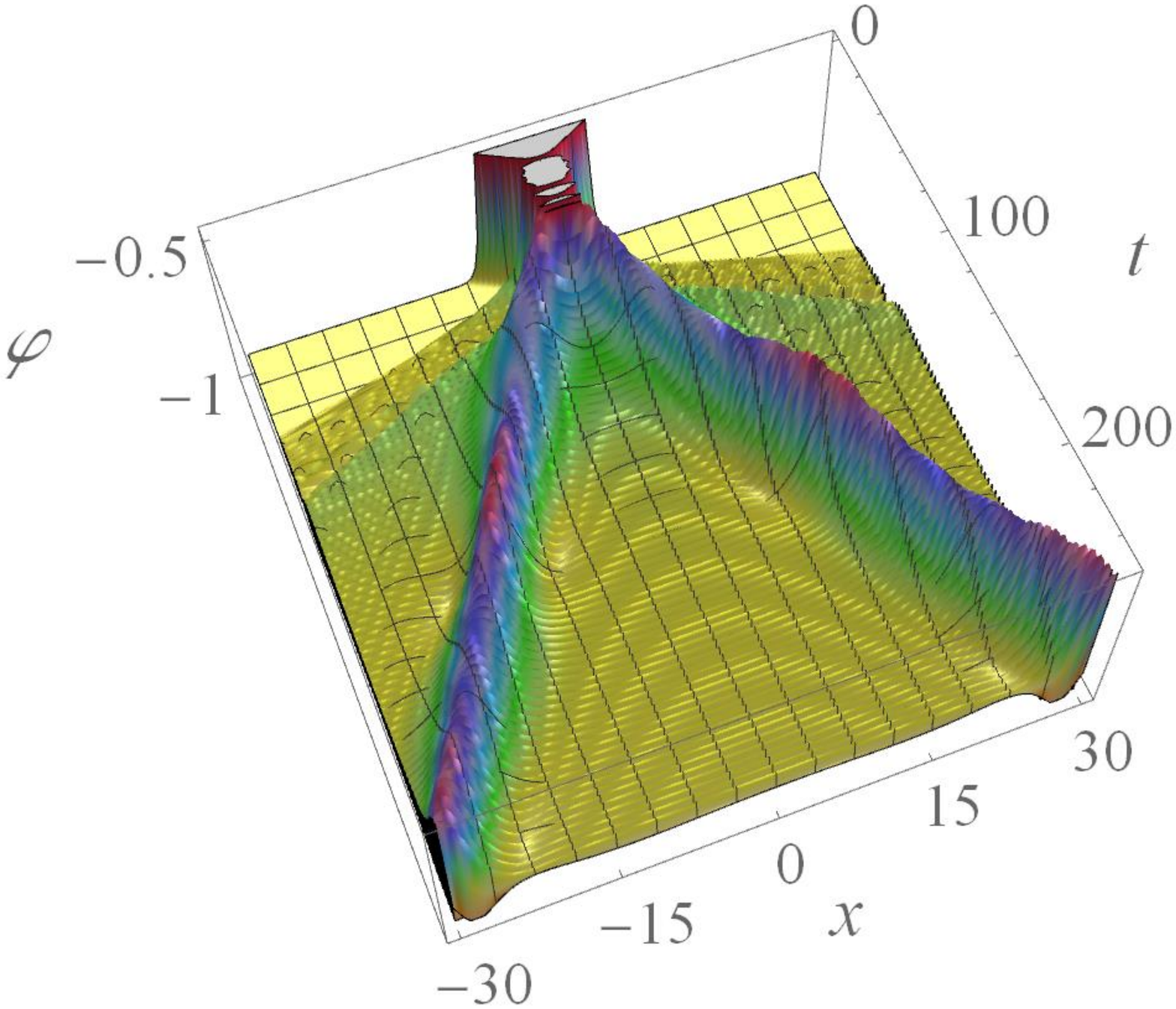}
\end{minipage}
}
\subfigure[$v_\mathrm{in}^{}=0.44207$, $v_\mathrm{os}^{}\approx 0.18$]{
\begin{minipage}{0.5\linewidth}
\centering\includegraphics[width=\linewidth]{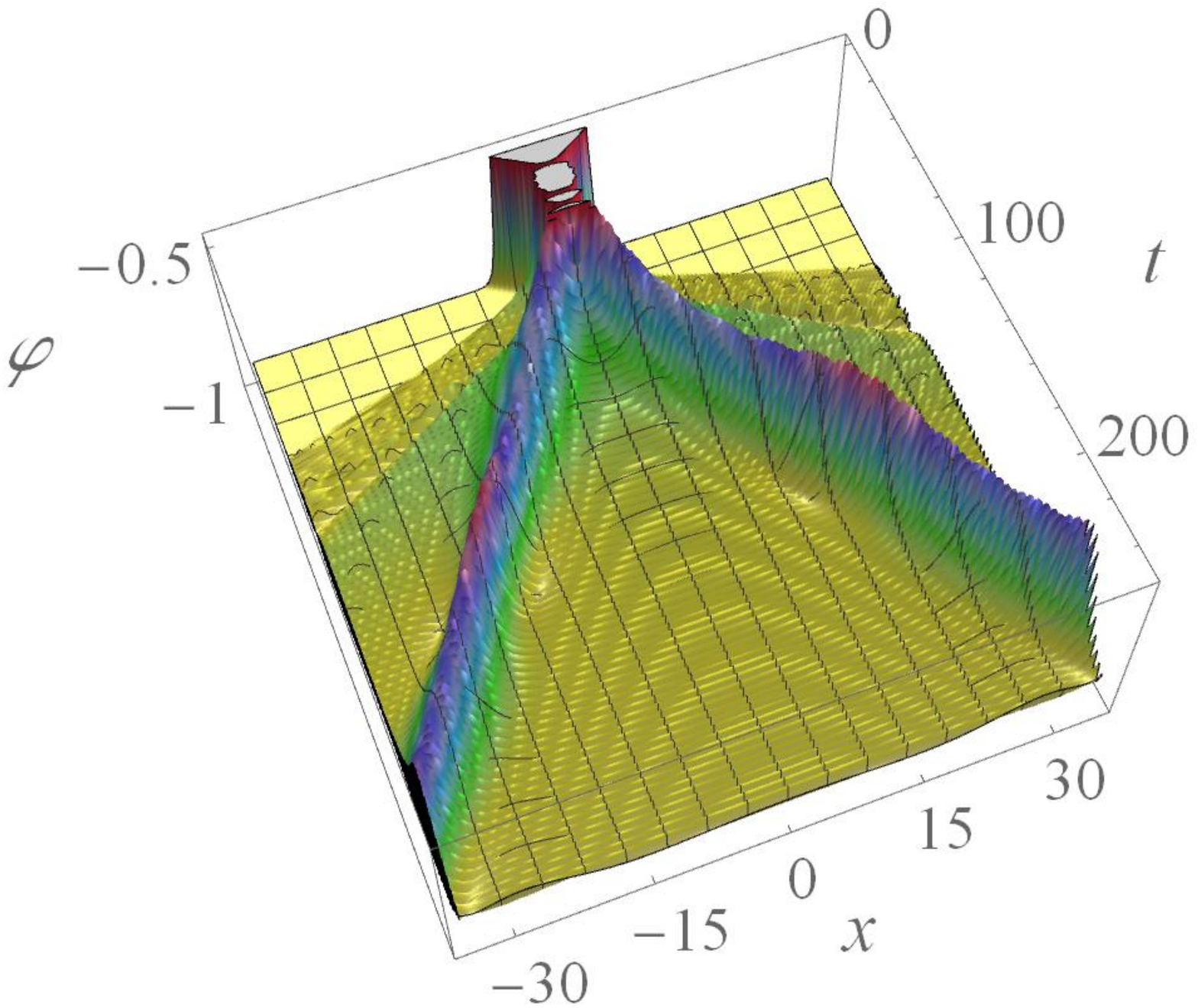}
\end{minipage}
}
\subfigure[$v_\mathrm{in}^{}=0.44220$, $v_\mathrm{os}^{}\approx 0.06$]{
\begin{minipage}{0.5\linewidth}
\centering\includegraphics[width=\linewidth]{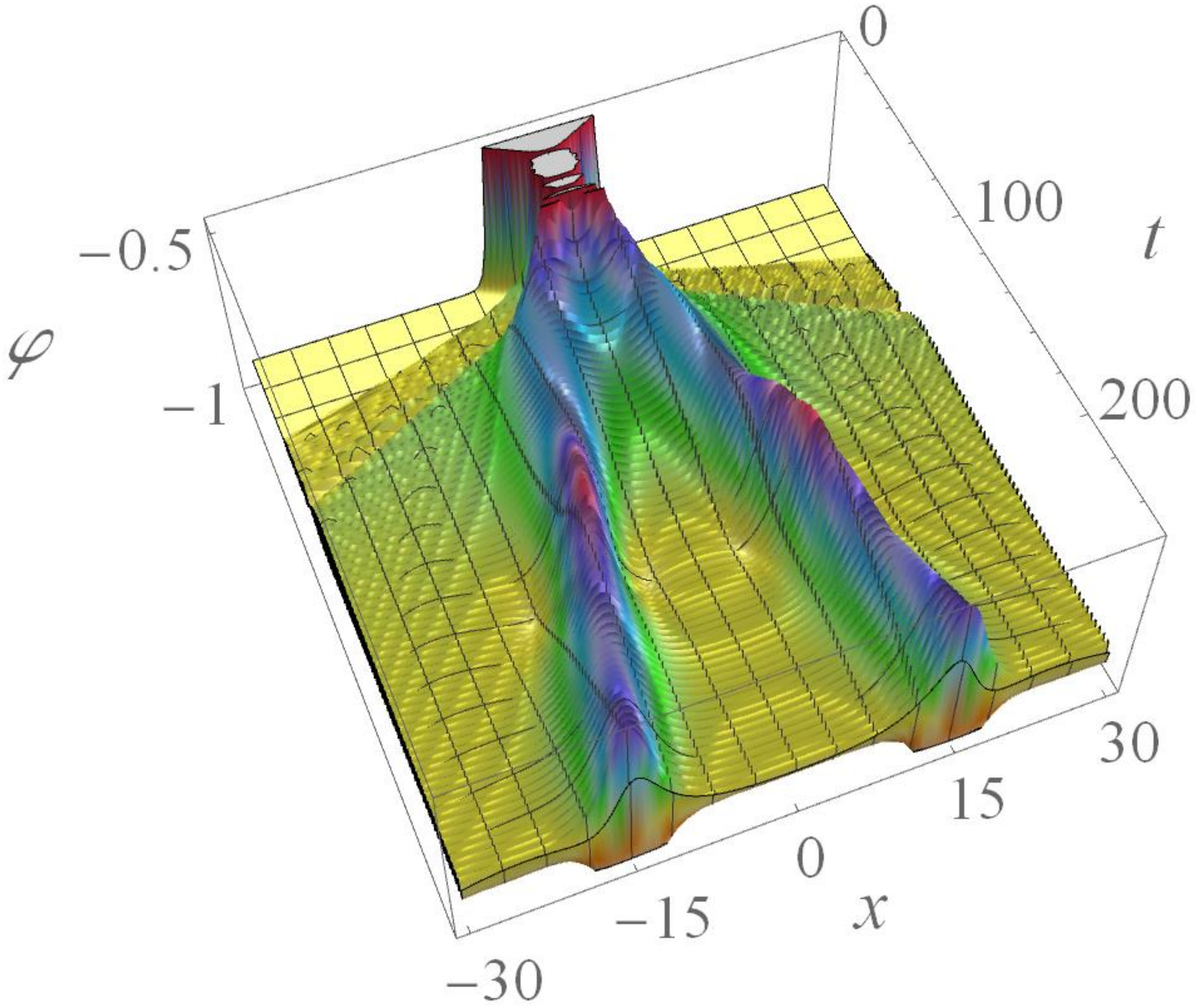}
\end{minipage}
}
\caption{Escape of two oscillons for several values of the initial velocity of the colliding kinks $v_\mathrm{in}^{}$, here $v_\mathrm{os}^{}$ stands for the final velocity of the escaping oscillon.}
\label{fig:bions_escape}
\end{figure}

At the same time, the bion formation in the range $v_\mathrm{in}^{}<v_\mathrm{cr}^{}$ outside the escape windows looks differently. In our numerical experiments we observed new phenomena, which are not typical for the $\varphi^4$ kinks. In many cases the final configuration looked like a bound state of two oscillons. These oscillons oscillate around each other near the origin and, as a consequence, the dependence on time of the field at the origin has a low-frequency envelope, as it is shown in figure \ref{fig:oscillating_bions_sinh_phi-4}.
One notes that the amplitude and frequency of oscillations of these structures depend on the initial velocity of the colliding kinks. Moreover, at some values of the initial velocity we observed escape of the two oscillons with the final velocity $v_\mathrm{os}^{}$, which varies in a wide range, as one can see from figure \ref{fig:bions_escape}. The situation can be interpreted as follows: at some initial velocities of the colliding kinks the bion is formed, which evolves rather fast into a bound state of two oscillons, which can either oscillate around each other, or escape to infinities. The intervals of the initial velocity of the colliding kinks, at which the oscillons escape, form {\it oscillons' escape windows}. The frequency of the field oscillations is the same for all oscillons, $\omega_\mathrm{os}^{}\approx 1.88$, which is very close to $\omega_1^{}=1.89$.

\begin{figure}[h!]
\vspace{-40ex}
\begin{minipage}{\linewidth}
\includegraphics[width=\linewidth]{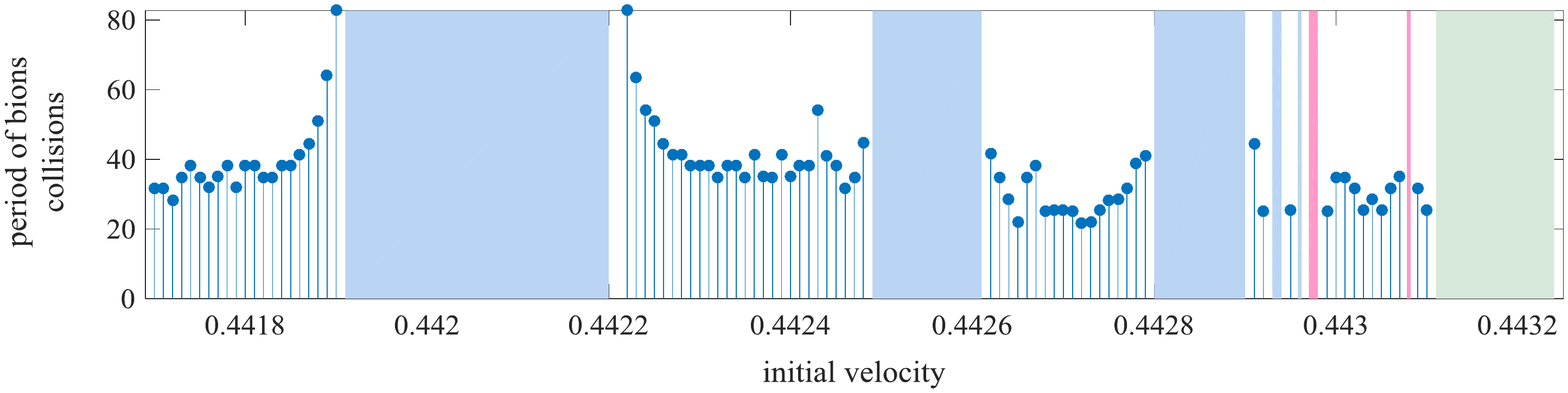}
\end{minipage}
\caption{The period of the oscillons' oscillations about each other as a function of the initial velocity of the colliding kinks. The blue-shaded areas denote the escape windows for oscillons. The green-shaded area denotes the two-bounce escape window for kinks, while the pink-shaded areas denote the four-bounce escape windows for kinks.}
\label{fig:period_bions_sinh_phi-4}
\end{figure}

In figure \ref{fig:period_bions_sinh_phi-4} we show the dependence of the period of oscillations on the initial velocity of the colliding kinks.
The shaded areas denote the escape windows for oscillons, i.e.\ the intervals of the initial velocity, at which the two oscillons escape to infinities. The widths of the escape windows are 0.00030, 0.00013, 0.00011, 0.00002, and $\sim 0.00001$ for the 1st, 2nd, 3rd, 4th, and 5th windows, respectively.

The high frequency in figure \ref{fig:oscillating_bions_sinh_phi-4} (right panels) is close to the frequency of the vibrational mode $\omega_1\approx 1.89$ of the sinh-deformed $\varphi^4$ kink. For example, at the initial velocity $v_\mathrm{in}^{}=0.44183$ the frequency is 1.86, at $v_\mathrm{in}^{}=0.44188$ it is 1.84, and at $v_\mathrm{in}^{}=0.44190$ it equals 1.83.

\section{Comments and Conclusion}
\label{sec:Conclusion}

In this work, we investigated the scattering of kinks of the sinh-deformed $\varphi^4$ model, obtained from the $\varphi^4$ model by the deformation procedure, and compared it with the same process in the $\varphi^4$ model. We showed that the two models engender similar behavior in several aspects: they support similar kinklike configurations, and their stability potentials present almost the same profile, which gives rise to the zero mode and the vibrational state with the frequency $\omega_1=\sqrt{3}\approx 1.73$ in the case of the $\varphi^4$ model, and $\omega_1^{}\approx 1.89$ for the sinh-deformed $\varphi^4$ model.

Moreover, in the scattering of kinks, the two models also admit a critical velocity $v_\mathrm{cr}^{}$, which separates two different regimes of the collisions. On the one hand, at $v_\mathrm{in}^{}<v_\mathrm{cr}^{}$ we observed the capture of kinks and the formation of bound states and, on the other hand, for $v_\mathrm{in}^{}>v_\mathrm{cr}^{}$ the kinks escape to infinity after one collision. The value of the critical velocity is $v_\mathrm{cr}^{}=0.4639$ for the sinh-deformed $\varphi^4$ model and for the $\varphi^4$ model it is equal to 0.2598.

In the study of collisions of kinks in the sinh-deformed $\varphi^4$ model, we observed that for velocities in the range $v_\mathrm{in}^{}<v_\mathrm{cr}^{}$, there appeared several escape windows, which are also specific for the $\varphi^4$ and some other models. In particular, we have found two-bounce, three-bounce, and four-bounce escape windows; recall that within an $n$-bounce window the kinks escape to infinities after $n$ collisions. The emergence of the escape windows is related to the resonant energy exchange between the translational and the vibrational modes of the kink and the antikink. 

The general results of the kink collisions in the sinh-deformed model suggest that the model is not integrable, and that its kinklike configuration is not a soliton. Interestingly, at this point one can make a connection with the  sine-Gordon model, which is an integrable model \cite{Book}. This model can also be obtained from the $\varphi^4$ model with the same deformation procedure: using the deformation function $f(\varphi)=\sin\varphi$, the potential $U_1(\varphi)$ in eq.~\eqref{eq:potential_phi-4} transforms into
\begin{equation}
U_3(\varphi)=\frac12 \cos^2\varphi.
\end{equation}
This is the potential of the sine-Gordon model, and its soliton solution can be written in the form
\begin{equation}
\varphi_\mathrm{s}^{}(x)=\arcsin(\tanh x).
\end{equation}
As is well-known, the corresponding stability potential supports the zero mode and no other bound state, and this helps one to understand its integrability. The sine-Gordon potential is periodic, in contrast to the $\varphi^4$ model described by a polynomial potential. In this sense, the sinh-deformed model which we have studied in this work seems to be farther away from the sine-Gordon and integrability, and hence it should present information that is absent in the $\varphi^4$ model.

With this motivation in mind, we then looked deeper into the escape windows in the sinh-deformed model and observed a new phenomenon, the conversion of the kink-antikink pair into a complex oscillating structure at the collision point at the origin. This structure can be interpreted as a bound state of two individual oscillons. It is interesting that at some initial velocities of the colliding kinks we observed the escape of these two oscillons.

The interval of initial velocities of the kinks, in which the kinks collide and form a bound state of two oscillons, which then escape, can be called an oscillons' escape window. In our simulations the final velocity of the escaping oscillons varies in a wide range from zero to $\sim 0.2$. Near the oscillons' escape window the period of the oscillons' oscillations in their bound state increases, see figure \ref{fig:period_bions_sinh_phi-4}.

As shown in the recent work \cite{Bazeia.IJMPA.2017}, we can introduce other models using the deformation function of the hyperbolic type. In particular, we can start with the $\varphi^6$ model studied before in \cite{lohe}, which supports no vibrational state. For instance, we can use the potential
\begin{equation}\label{varphi6}
U_4(\varphi)=\frac{1}{2}\varphi^2 (1-\varphi^2)^2,
\end{equation}
and the deformation function $f(\varphi)=\sinh\varphi$ to get to a new model
\begin{equation}\label{varphi6d}
U_5(\varphi)=\frac{1}{2}\tanh^2\varphi\:(1-\sinh^2\varphi)^2.
\end{equation}
We note that for $\varphi$ very small, the above model \eqref{varphi6d} leads us back to the model in \eqref{varphi6}. We call this model \eqref{varphi6d} the sinh-deformed $\varphi^6$ model. It would be interesting to study the scattering of kinks in this model, to see how it can be connected to the investigation \cite{lohe,GaKuLi,MGSDJ,dorey}, which revealed a resonant scattering structure that provided a counterexample to the belief that the existence of the vibrational bound state is a necessary condition for the appearance of multibounce resonances. Another issue is the study of the force between two kinks, to see if it can be connected with the scattering of kinks.

We can also consider models with modified kinematics, as the ones recently investigated in \cite{DBI}, where one considers the Dirac-Born-Infeld case. This modification changes the standard scenario and may contribute to add new possibilities to the escape windows that appear in the standard situation. Another route concerns models described by two real scalar fields, as the one investigated in refs.~\cite{Bazeia.PLA.1995,Bazeia.PRD.2000}. In this case, the presence of the two fields leads to analytical kinklike solutions whose internal structure can be used to model Bloch walls. The scenario here is richer, and the study of the kinks scattering in this model would allow one to see how the internal structure contributes to the presence of the escape windows, etc. These and other similar issues are currently under consideration, and we hope to report on them in the near future.

\vspace{10mm}

\section*{Acknowledgments}

The authors would like to thank Dr.~Vadim Lensky for reading the manuscript and for valuable comments. This work was performed using resources of the NRNU MEPhI high-performance computing center. The research was supported by the Brazilian agency CNPq under contracts 455931/2014-3 and 306614/2014-6, and by the MEPhI Academic Excellence Project under contract No.~02.a03.21.0005, 27.08.2013.


\end{document}